\documentclass[sigconf]{acmart}
\usepackage{subcaption}
\usepackage{balance}
\setcopyright{acmcopyright}
\usepackage{multirow}
\newcommand{\modelname}[1]{RTM}
\usepackage{array}
\newcolumntype{C}[1]{>{\centering\arraybackslash}m{#1}}
\newcolumntype{R}[1]{>{\rtab\arraybackslash}m{#1}}

\AtBeginDocument{%
	\providecommand\BibTeX{{%
			\normalfont B\kern-0.5em{\scshape i\kern-0.25em b}\kern-0.8em\TeX}}}

\copyrightyear{2021} 
\acmYear{2021} 
\setcopyright{acmcopyright}\acmConference[SIGIR '21]{Proceedings of the 44th International ACM SIGIR Conference on Research and Development in Information Retrieval}{July 11--15, 2021}{Virtual Event, Canada}
\acmBooktitle{Proceedings of the 44th International ACM SIGIR Conference on Research and Development in Information Retrieval (SIGIR '21), July 11--15, 2021, Virtual Event, Canada}
\acmPrice{15.00}
\acmDOI{10.1145/3404835.3462911}
\acmISBN{978-1-4503-8037-9/21/07}

\begin{document}
	\fancyhead{}
	
%	\title{A Review-based Transformer Model for Personalized Product Search}
	\title{Learning a Fine-Grained Review-based Transformer Model for Personalized Product Search}
	% The indent in the affliation block matters
	\author{Keping Bi}
	\affiliation{
		\institution{University of Massachusetts Amherst}
		\city{Amherst} 
		\state{MA} 
		\country{USA}
		%	\postcode{01003-9264}
	}
	\email{kbi@cs.umass.edu}

	\author{Qingyao Ai}
	\affiliation{
		\institution{University of Utah}
		\city{Salt Lake City} 
		\state{UT} 
		\country{USA}
	%	\postcode{01003-9264}
	}
	\email{aiqy@cs.utah.edu}

	\author{W. Bruce Croft}
	\affiliation{%
		\institution{University of Massachusetts Amherst}
		\city{Amherst} 
		\state{MA} 
		\country{USA}
	%	\postcode{01003-9264}
	}
	\email{croft@cs.umass.edu}

% 	\author{Ben Trovato}
% 	\authornote{Both authors contributed equally to this research.}
% 	\email{trovato@corporation.com}
% 	\orcid{1234-5678-9012}
% 	\author{G.K.M. Tobin}
% 	\authornotemark[1]
% 	\email{webmaster@marysville-ohio.com}
% 	\affiliation{%
% 		\institution{Institute for Clarity in Documentation}
% 		\streetaddress{P.O. Box 1212}
% 		\city{Dublin}
% 		\state{Ohio}
% 		\postcode{43017-6221}
% 	}
	
	\renewcommand{\shortauthors}{Trovato and Tobin, et al.}
%	\balance
	\begin{abstract}
		%In product search, customers make purchase decisions based on not only the product relevance but also their personal preferences. Despite its great potential, recent analysis on commercial logs shows that personalization does not always improve the quality of product search. Most existing personalized product retrieval models, however, cannot control whether to apply personalization or not under different contexts. 
		%From the perspective of model design, 
		Product search has been a crucial entry point to serve people shopping online. 
		Most existing personalized product models follow the paradigm of representing and matching user intents and items in the semantic space, where finer-grained matching is totally discarded and the ranking of an item cannot be explained further than just user/item level similarity. In addition, while some models in existing studies have created dynamic user representations based on search context, their representations for items are static across all search sessions. This makes every piece of information about the item always equally important in representing the item during matching with various user intents.
		%Also, users and items have their unique vectors according to their identifiers most of the time, making the models difficult to generalize to unseen users and items. 
		Aware of the above limitations, we propose a review-based transformer model (\modelname{}) for personalized product search, which encodes the sequence of query, user reviews, and item reviews with a transformer architecture. 
		%\modelname{} can perform adaptive personalization with different accumulative attention weights on user reviews in various search sessions. 
		\modelname{} conducts review-level matching between the user and item, where each review has a dynamic effect according to the context in the sequence. This makes it possible to identify useful reviews to explain the scoring. 
		%Experimental results show that \modelname{} achieves significantly better performances than state-of-the-art personalized product search baselines. 
		Experimental results show that \modelname{} significantly outperforms state-of-the-art personalized product search baselines. 
		
	\end{abstract}
	

	\begin{comment}
	
	\begin{CCSXML}
	<ccs2012>
	<concept>
	<concept_id>10002951.10003317.10003331.10003271</concept_id>
	<concept_desc>Information systems~Personalization</concept_desc>
	<concept_significance>500</concept_significance>
	</concept>
	<concept>
	<concept_id>10002951.10003317.10003338</concept_id>
	<concept_desc>Information systems~Retrieval models and ranking</concept_desc>
	<concept_significance>500</concept_significance>
	</concept>
	<concept>
	<concept_id>10002951.10003317.10003371</concept_id>
	<concept_desc>Information systems~Specialized information retrieval</concept_desc>
	<concept_significance>500</concept_significance>
	</concept>
	</ccs2012>
	\end{CCSXML}
	
	\ccsdesc[500]{Information systems~Personalization}
	\ccsdesc[500]{Information systems~Retrieval models and ranking}
	\ccsdesc[500]{Information systems~Specialized information retrieval}
	
	\end{comment}
	
	\keywords{Product Search; Personalization; Transformer Models; Fine-grained Matching; Review-based Matching}
	
	\maketitle

\section{Introduction}
\label{sec:introduction}

In product search, users' purchase behaviors usually depend on their individual preferences in addition to product relevance. Aware of this point, recent studies \cite{ai2017learning, ai2019explainable,bi2020transformer,zou2019learning, xiao2019dynamic, guo2019attentive} have explored to incorporate personalization in the product retrieval models and produced significant improvements in the search quality. 
A typical paradigm of existing personalized product search models is to represent the user intents and items explicitly with embedding vectors and match them in the latent space with dot product or cosine similarity to yield the item score \cite{ai2017learning, ai2019explainable, guo2019attentive,bi2019conversational,ai2019zero, bi2020transformer}. 
Under this paradigm, user's search intents are usually represented by a function of the query vector and the user vector, which can be a convex combination \cite{ai2017learning}, a simple addition \cite{ai2019zero}, several neural layers \cite{guo2019attentive}, or a transformer encoder \cite{bi2020transformer}. Item representations are usually learned by predicting words in the item reviews \cite{ai2017learning,ai2019explainable,ai2019zero,guo2019attentive, bi2019conversational, bi2020transformer}, and user vectors, which represents the user's personalized preferences, are represented similarly \cite{ai2017learning} or as a weighted combination of items vectors with attention mechanisms \cite{ai2019zero,guo2019attentive, bi2020transformer}. 

Despite its popularity, the existing product search paradigm has several limitations in practice.
%\textbf{Lack of Finer-grained Matching and Explanation}. 
First, in the existing product search framework, the item scoring is based on matching at the user/item level \cite{ai2017learning, ai2019zero, guo2019attentive, bi2020transformer} instead of the finer-grained level, e.g., user/item reviews. Thus how a specific user preference mentioned in the user reviews matches an item property indicated in the item reviews could not be captured sufficiently. While a top-retrieved item is close to the user intent or some of the users' historical purchases in the latent space, why the model considers them close is not clear. 
%In this case, we only know a top-retrieved item is close to the user intent in the latent space, or we may also know which previously purchased items by the user the retrieved item is related to the most. The reason why the model considers them close or related in the semantic space is not clear. 
Second, 
%\textbf{Static Item Representation}. 
despite their efforts on constructing dynamic user representations under different contexts \cite{ai2019zero,guo2019attentive, bi2020transformer}, existing product search studies always represent items statically regardless of the context  \cite{ai2017learning, ai2019explainable, guo2019attentive,bi2019conversational,ai2019zero, bi2020transformer}. During the representation learning, all the associated reviews are considered to have equal importance. However, the same aspect of an item may play different roles in representing the item when matching with various user preferences towards the aspect. For example, for a user who has tooth whitening needs, reviews that comment on the toothpaste's whitening function should be more important to represent the item than the long-lasting breath freshening property. 
However, in existing product search models, a review complaining about the shipment and package handling of the toothpaste may be considered equally important as other reviews since the item representations are built as static vectors without considering user intents.
%for a user who cares about fresh breath, this latter property should be emphasized more than the whitening function. 
%Thus, there are potential benefits to dynamically weigh each piece of information about the item during matching with user intents. 
\begin{comment}
3) \textbf{Inferior Generalization Ability}. Under this paradigm, most models construct unique user and item representations according to user and item identifiers beforehand, which means that they cannot generalize to users and items that haven't appeared in the training process even though these users and items get several associated reviews after some time. This can be detrimental to new products and customers who have some reviews but their identifiers are not included in the training set. 
\end{comment}

% This, however, is problematic as the scale of e-commerce platforms overgrows with new products and new customers every day.
% The paradigm requires unique representation for each user and item according to their identifiers only with one exception in \cite{ai2019zero} that represents users with a weighted combination of the vectors of purchased items without building user-specific lookup tables. This makes the models unable to generalize to users and items that have not appeared in the training set, which is detrimental to newly released products and new customers even when they have a few associated reviews. 

Given these limitations, in this paper, we propose to match user intents and items at the level of finer-grained information (e.g., their associated reviews) instead of explicitly representing them with static vectors. Specifically, we score an item based on the sequence of the query, user reviews, and item reviews with a transformer \cite{vaswani2017attention} architecture, where every unit, i.e., query or user/item review, could interact with each other during matching. We refer to our model as the review-based transformer model, abbreviated as \modelname{}.  

\modelname{} has a couple of advantages over existing Transformer-based \cite{bi2020transformer} or other neural model based product search approaches \cite{ai2017learning, ai2019zero,guo2019attentive}. %,  \modelname{} has a couple of advantages.
First, \modelname{} conducts user-item matching at the review level so that the reason an item is ranked to the top can be explained by some useful reviews that draw more attention from other units in the sequence. 
Second, in \modelname{}, the importance of each user and item review during matching is dynamically adapted in different sequences, where both users and items carry dynamic information under different context. When encoded by a multi-layer \modelname{}, the review representation is dynamically changed according to its interactions with other units in the sequence. 
Also, \modelname{} represents user and item based only on their reviews without the need for their identifiers, so it can easily generalize to the users and items that have associated reviews but have not appeared in the training set. 
% Moreover, based on the analysis of commercial search logs, \citet{ai2019zero} have found that personalization has a positive effect only when the query carries diverse purchase intents and the individual preferences significantly differ from the group preference under a query. Therefore, to maximize the benefit of personalization across all search sessions, it is essential to identify when and when not to apply personalization. 
Last but not least, 
%\citet{ai2019zero} have found that it is essential to perform adaptive personalization in different search sessions since universal personalization can harm some search sessions. 
% In this regard, our 
\modelname{} can conduct more flexible personalization than most existing personalized product search models  \cite{ai2017learning,ai2019explainable,guo2019attentive, zhang2018towards, bi2019conversational}. Personalization in \modelname{} can vary from no to full effect depending on the contexts since the user reviews could have zero accumulative attention weights and so does the query. 
Our experimental results confirm our model's advantages by showing that \modelname{} significantly outperforms the state-of-art baselines. To better understand the model and explain the search results, we also analyze the influence of different settings on \modelname{} and conduct case studies to show how \modelname{} identifies important information during matching.

Our contributions in this paper can be summarized as follows: 
\begin{comment}
\begin{itemize}
    \item We propose a review-based transformer model (\modelname{}) for personalized product search that is superior to existing models in terms of finer-grained matching, dynamic user/item representation, generalization ability, and the influence of personalization;
    \item Our experimental results show that our \modelname{} achieves significantly better performance than state-of-the-art personalized product search techniques; 
    \item We analyze the model property and conduct case studies to understand the model behaviors better.
\end{itemize}
\end{comment}
1) we propose a review-based transformer model (\modelname{}) for personalized product search that is superior to existing models in terms of finer-grained matching, dynamic user/item representation, generalization ability, and the influence of personalization; 2) our experimental results show that our \modelname{} achieves significantly better performance than state-of-the-art personalized product search techniques; 3) we analyze the model property and conduct case studies to understand the model behaviors better.
\begin{comment}
\end{comment}
%3) we analyze the influence of different settings on \modelname{} and conduct case studies to show how \modelname{} identifies important information during matching, which helps us to understand the model better and explain the search results. 

	%!TEX root=main.tex
\section{Related Work}
\label{sec:related_work}
Our work is closely related to product search, personalized web search, and transformer-based retrieval models. 
%\subsection{Product Search}

\textbf{Product Search.}
Since product information is more structured, earlier research uses facets such as brands, prices, and categories for product search \cite{lim2010multi,vandic2013facet}. However, these approaches cannot handle free-form user queries. To support search based on keyword queries, \citet{duan2013probabilistic,duan2015mining} extended language-model-based techniques by assuming that queries are generated from the mixture of one language model of the background corpus and the other one of products conditioned on their specifications. Word mismatch problems still exist in these approaches. \citet{van2016learning} noticed this problem and proposed representing and matching queries and products in the latent semantic space. 

% \citet{ai2017learning} realized the importance of personalization on product search and proposed a hierarchical embedding model where the representations of users and items are learned by predicting words in their associated reviews, and a convex combination of the query and user vector is used to predict purchased items. \citet{guo2019attentive} represent long and short-term user preferences with an attention mechanism applied to users' recent purchases and their global vectors. Recently, \citet{ai2019zero} observed that personalization does not always have a positive effect from the analysis of commercial search logs. They further proposed a zero-attention model that can control the influence of personalization by allowing the zero vector to have various attention weights. 

Aware of the importance of personalization in product search,
\citet{ai2017learning} proposed a hierarchical embedding model where they use a convex combination of the query and user vector to predict purchased items. \citet{guo2019attentive} represent long and short-term user preferences with an attention mechanism applied to users' recent purchases and their global vectors. Recently, from the analysis of commercial search logs, \citet{ai2019zero} observed that personalization does not always have a positive effect. They further proposed a zero-attention model (ZAM) that can control the influence of personalization. However, the maximal effect personalization can have is equal to the query. \citet{bi2020transformer} found this limitation and proposed a transformer model to encode the query and historically purchased items where personalization can have none to full effect. 
% by allowing the zero vector to have various attention weights. 

There are also studies on other aspects such as popularity, other information sources (e.g., images), diversity, and labels for training in product search. \citet{long2012enhancing} incorporated popularity with relevance for product ranking. \citet{di2014relevance} and \citet{guo2018multi} investigated on using images as a complementary signal. \citet{ai2019explainable} proposed an explainable product search model with dynamic relations such as brand, category, also-viewed, also-bought, etc. Efforts have also been made to improve result diversity to satisfy different user intents behind the same query \cite{parikh2011beyond,yu2014latent}. In terms of training signals, \citet{wu2018turning} jointly modeled clicks and purchases in a learning-to-rank framework and \cite{karmaker2017application} compared the effect of different labels such as click-rate, add-to-cart ratios, and order rates. More recently, \citet{zhang2019neural} integrated the graph-based feature with neural retrieval models for product search. \citet{xiao2019dynamic} studied personalized product search under streaming scenarios.  \citet{ahuja2020language} learned language-agnostic representations for queries and items that can support search with multiple languages. There are also studies on interactive product search such as in a multi-page search setting \cite{bi2019study} and in conversational systems \cite{zhang2018towards,bi2019conversational}.

Most existing work either focuses on non-personalized product search or conducts personalized product search with static user/item representations. In contrast, we propose an adaptive personalization model for product search and conduct item scoring in a novel paradigm, i.e., dynamic matching user intents and items at the review level instead of explicitly represent them in the semantic space and match them directly. 

%\subsection{Personalized Web Search}
\textbf{Personalized Web Search.}
Personalization has also been studied in the context of Web search, where results are customized to each user in order to maximize their satisfaction \cite{dumais2016personalized}. Existing approaches usually infer users' personal needs from their locations, search histories, clicked documents, etc., and then re-rank results accordingly \cite{bennett2012modeling, chirita2005using,shen2005implicit,teevan2005personalizing,cheng2021long}. While users could behave differently for the same query \cite{dou2007large,white2007investigating}, personalization does not always benefit search quality. Based on the analysis of user behavior patterns in the large-scale logs on Live Search, \citet{teevan2008personalize} observed that the effectiveness of personalization in Web search depends on multiple factors such as result entropy, result quality, search tasks, and so on. 

We focus on product search in this paper, where personalization is more appealing than in Web search. While relevance is usually the primary criterion in Web search, user purchases depend on both item relevance and user preferences in product search. 

%\subsection{Transformer-based Retrieval Models}
\textbf{Transformer-based Retrieval Models.}
% There has been sparse research using transformers for retrieval. 
After the pre-trained contextual language models, i.e., BERT \cite{devlin2018bert}, grounded on the transformer architecture, achieved compelling performance on a wide range of natural language processing tasks, more studies have explored leveraging BERT in information retrieval tasks as well. \citet{nogueira2019passage} show BERT's effectiveness on passage ranking, and \citet{dai2019deeper} demonstrate that BERT can leverage language structures better and enhance retrieval performance on queries in natural languages. \citet{wuleveraging} proposed a passage cumulative gain model that applies a sequential encoding layer on top of the BERT output of a query-passage pair to score a document. \citet{qu2019attentive} refine the original BERT model with an additional attention layer on each question-utterance pair to attentively select useful history to identify the answer span in conversational question answering. 

Our model is based on transformers instead of BERT. In other words, we leverage the transformer architecture to dynamically match the query-user pair and the item at the review level but we do not represent queries and reviews with pre-trained BERT.
%\citet{camaradiagnosing}

	%!TEX root=main.tex

\begin{comment}
Model:
Query/review representation. 
Better than pv-based method. 
Personalized Item Scoring:
Encoding matching sequence. 
position embeddings (Sequence matters: recent purchases play different roles from long-ago purchases)
segment embeddings: differentiate user reviews and item reviews. 
addible components
position embeddings may also indicate the user and item reviews. 
q^{(l)}
score the sequence. 
map to score
Q-U-Item interaction. 
Interpretation:
No explicit user/item representation. 
User's purchase intent
self-attention mechanism given query. 
personalization: 
from 0 to 1. 
dynamically changed according to items. different from ZAM, TEM
review-based compared with item-based
care about specific properties of the item,  rather than aggregating all the information of the item. 
Explanable Item Matching. 
Which reviews attract more attention. 
%Dynamic item representation based on reviews 
Dynamic Review Representation:
intent -> intent:     review representation dynamically changed
ru_a-> ru_b 
item review -> item review
matching of each review in terms of various previous interest. 
how is each review as an interest satisfied with the item (all the associated reviews.)
Comparison with TEM (a paragraph)
Single layer:
Multiple layer:
Optimization Framework:
P(i|u, q) = softmaxf(q,u,i)

\end{comment}

\section{Review-based Transformer Model}
\label{sec:review_transformer}
%Beginning paragraph
This section introduces each component of our review-based transformer model (RTM) and how RTM conducts personalized item scoring. Then we show how RTM is optimized and provide interpretations of RTM from the perspective of model design.  

\begin{figure}
	\centering
	\includegraphics[width=0.48\textwidth]{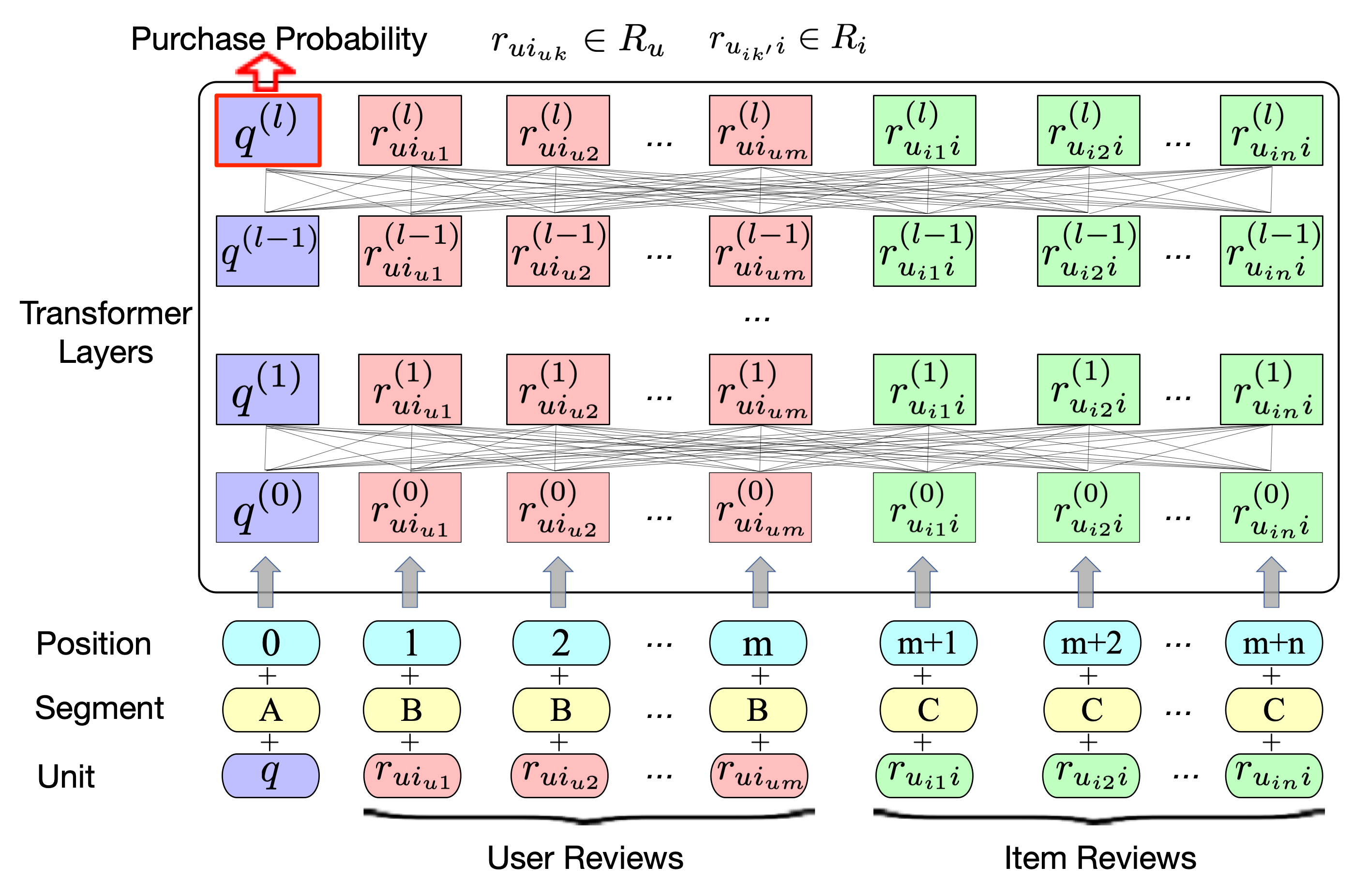} %pdf,png
	\caption{Our Review-based Transformer Model (RTM). }
	%$I_u$ is the historical purchase of user $u$; $R_i$ represents the reviews of item $i$; $q$ and $i_uj$ represent query and the $j$-th item $u$ purchases. 
	\label{fig:rtm}
\end{figure}
\subsection{Personalized Item Scoring}
\label{subsec:personal_score}
In contrast to previous studies that have explicit representations for both users and items \cite{ai2017learning, ai2019zero, guo2019attentive} or just for items \cite{ai2019zero, guo2019attentive, bi2020transformer}, our model does not represent the user or item with a single vector. Instead, we consider the user and item's historical reviews as the basic unit carrying their information and learn to score the item given the query-user pair using the interactions between their basic information units. In this way, the matching of a query-user pair and an item can be conducted at a finer grain and could potentially capture more connections between the user and the item, which we will illustrate later in Section \ref{subsec:model_interpretation} in detail. 

Let $q$ be a query submitted by a user $u$ and $i$ be an item in $S_i$, which is the collection of all the items. $R_u=(r_{ui_{u1}},r_{ui_{u2}}, \cdots,  r_{ui_{um}})$ and $R_i=(r_{u_{i1}i}, r_{u_{i2}i}, \cdots, r_{u_{in}i})$ denote the sequence of $m$ and $n$ reviews associated with $u$ and $i$ respectively in a chronological order, where $r_{ui_{uk}} (1 \leq k\leq m)$ is the review $u$ wrote for her $k$-th purchased item $i_{uk}$ and $r_{u_{ik'}i} (1 \leq k'\leq n)$ is the review associated with $u_{ik'}$, which is the $k'$-th user who purchased $i$. $m$ and $n$ are the length of $R_u$ and $R_i$ respectively. As shown in Figure \ref{fig:rtm}, we feed the sequence of $(q, R_u, R_i)$ to an $l-$layer transformer encoder to let the query and user's purchase history interact with the item's associated reviews.

\textbf{Input Embeddings.}
Inspired by the architecture of BERT \cite{devlin2018bert}, our model also has three types of embeddings associated with each input unit (query or review) to the transformer encoder. They are the \textit{unit embedding} that is computed from the words in the unit, which we will introduce later in Section \ref{subsec:query_vec}, the \textit{position embeddings} \cite{vaswani2017attention} that indicate the position of a unit in the sequence, and the \textit{segment embeddings} that differentiate whether the unit is the query $q$, a review from $u$, or a review about $i$, denoted as $A,B$, and $C$ respectively, as shown in Figure \ref{fig:rtm}.

Since a recent review from the user may indicate her current intention better than her long-ago reviews and an item's recent review may reveal the item's current properties better than an old review about the item, position embeddings could be beneficial by indicating the temporal order of each review. While the time of reviews could be significant in some categories, it is also possible that the information of items such as the music on a CD is static, and user preferences behind particular purchase needs stay similar for a long time. Therefore, we make the position embeddings an optional choice in our model. 

The segment embedding of a unit is also optional in \modelname{} since the model could infer which segment the current unit belongs to with their position embeddings. The reviews about $i$ always have later positions than reviews of $u$, and $q$ is always at position 0. When position embeddings are not necessary in some cases, segment embeddings are still needed to differentiate the input units. We will show the effects of position and segment embeddings on our model later in Section \ref{subsec:model_analysis}.  

Formally, the input vector of each unit is the sum of the unit embedding and its associated optional position embedding and segment embedding:
\begin{equation}
\label{eq:q0}
\begin{split} %same as aligned
%\begin{aligned}
\mathbf{q}^{(0)} &= \mathbf{q} + \mathcal{I}_{pos} E(0) + \mathcal{I}_{seg} E(A) \\
\mathbf{r_{ui_{uk}}}^{(0)} &= \mathbf{r_{ui_{uk}}} + \mathcal{I}_{pos} E(k) + \mathcal{I}_{seg} E(B), 1\leq k\leq m \\
\mathbf{r_{u_{ik'}i}}^{(0)} &= \mathbf{r_{u_{ik'}i}} + \mathcal{I}_{pos} E(m+k') + \mathcal{I}_{seg} E(C), 1\leq k' \leq n \\
%\end{aligned}
\end{split}
\end{equation}
where $E(\cdot)$ is the embedding of position $0, 1, \cdots, m+n$ or segment $A, B, C$; $\mathcal{I}_{pos} \in \{0,1\}$ and $\mathcal{I}_{seg} \in \{0,1\}$ indicate whether position and segment embeddings are used in the computation; $\mathbf{q}, \mathbf{r_{ui_{uk}}}$, and $\mathbf{r_{u_{ik'}i}}$ are the vector representation of $q, r_{ui_{uk}}$, and $r_{u_{ik'}i}$ respectively, which we will introduce in Section \ref{subsec:query_vec}. 

\textbf{Transformer Layers.}
The input sequence of vectors is then passing through transformer layers where the units interact with each other. As in \cite{vaswani2017attention}, each transformer layer has two sub-layers: a multi-head self-attention mechanism and a position-wise fully connected feed-forward network. 

Let $K$ and $V$ denote vectors of the sequence $(q, r_{ui_{u1}}, \cdots,  r_{ui_{um}}$, 
$r_{u_{i1}i}, \cdots, r_{u_{in}i})$ and let $Q$ be the vector of any unit in the sequence. For attention head $j$, the output vector is computed as a weighted sum of values $V$ according to the attention their corresponding keys $K$ obtain with respect to $Q$, i.e., 
\begin{equation}
\label{eq:attnhead}
\text{Attn}_j(Q,K,V) = \text{softmax} (\frac{Q_jK_j^T}{\sqrt{d/h}})V_j
\end{equation}
where $d$ is the dimension size of a unit vector and $h$ is the number of attention heads. $Q_j=QW_j^Q, K_j=KW_j^K, V_j=VW_j^V$ and $W_j, K_j, V_j$ are project matrices for Attn$_j$. 
Attn is applied for $h$ attention heads, and each output is concatenated and mapped to the final yield of the multiple-head attention (MultiHeadAttn).

The position-wise feed-forward network (FFN) applies the same transformation to each position separately and identically. 
Then there is also a residual connection for each sub-layer, which allows the input to go through the layer directly, followed by layer normalization. 
So the output vector $x^{(t)}$ in the $t$-th transformer layer $(1\leq t\leq l)$ at an arbitrary position from $0$ to $m+n$ can be computed as:
\begin{equation}
\begin{aligned}
\label{eq:transforme_layer}
&x^{(t)} = \text{LayerNorm}(y+\text{FFN}(y)) \\
&y = \text{LayerNorm}(x^{(t-1)}+\text{MultiHeadAttn}(x^{(t-1)},K,V))
\end{aligned}
\end{equation}
where $x^{(t-1)}$ is the output of the $(t-1)$-th layer, which can be obtained with Eq. \ref{eq:transforme_layer} when $t>1$ and Eq. \ref{eq:q0} when $t=1$. For more details about the transformers, please refer to \cite{vaswani2017attention}. 

\textbf{Final Score.}
At last, we use the output query vector at the final layer, i.e., $\mathbf{q^{(l)}}$, which involves all the interactions between every unit in the sequence with query $q$ to score item $i$. Specifically, the score of $i$ given $q$ and $u$ is computed with function $f$:
\begin{equation}
\label{eq:item_score}
f(q,u,i) = \mathbf{q^{(l)}} W_o 
\end{equation} 
where $W_o \in \mathbb{R}^{d \times 1}$. 
%We choose $\mathbf{q^{(l)}}$ instead of vectors at other positions because it is computed by attending to each unit in the sequence using query $q$, which is more reasonable than using any of the reviews.

\modelname{} can be degraded to a non-personalized model when there are no user reviews available. Also, \modelname{} can score an item based on its descriptions when it does not have associated reviews. 

\subsection{Query/Review Representation}
\label{subsec:query_vec}
It is important to compute query and review representations on the fly so that the model can handle arbitrary queries and reviews during inference time. 
%Since queries are usually unknown in advance, it is important that their representation can be computed on the fly during inference time so that arbitrary queries can be handled. 
Previous studies \cite{ai2017learning, van2016learning} have explored methods to construct query embeddings directly from query words, such as averaging word embeddings or applying recurrent neural networks on the query word sequence. A state-of-the-art technique is to encode a query with a non-linear projection $\phi$ on the average query word vectors:
\begin{equation}
\label{eq:qvec}
\mathbf{q} = \phi(\{w_q| w_q \in q\}) = \tanh(W_{\phi_q} \cdot \frac{\sum_{w_q\in q}\mathbf{w_q}}{|q|} +b_{\phi_q})
\end{equation}
where $W_{\phi_q} \in \mathbb{R}^{d \times d}$ and $b_{\phi_q} \in \mathbb{R}^{d}$ are learned during training,  $|q|$ is the length of query $q$, and $\mathbf{w_q}\in\mathbb{R}^d$ is the embedding of word $w_q$ in $q$. We use the same projection function $\phi$ with two different parameters $W_{\phi_r}$ and $b_{\phi_r}$ to collect the initial input representation of review $r$. 

In this way, word embeddings are shared across reviews toward the ranking optimization goal; significant words in the reviews can be emphasized with more weights in the matrix $W_{\phi_r}$. Thus the interaction between reviews can capture their keyword matching. 

%
%To collect the embedding of review $r$, we use the same projection function $\phi$ with two different parameters $W_{\phi_r}$ and $b_{\phi_r}$ to encode words in $r$:
%\begin{equation}
%\label{eq:rvec}
%\mathbf{r} = \phi(\{w_r| w_r \in r\}) = \tanh(W_{\phi_r} \cdot \frac{\sum_{w_r\in r}\mathbf{w_r}}{|r|} +b_{\phi_r})
%\end{equation}
We also considered representing reviews with another popular method, i.e., paragraph vectors \cite{le2014distributed}, by predicting words in the review with the review vector. However, paragraph vectors need to be trained beforehand and thus are difficult to generalize to unseen reviews. 
Moreover, in paragraph vectors, word embeddings can only be updated by predicting the words with these review vectors, which is an unsupervised signal regardless of user purchases. 
%In contrast, word embeddings can be optimized towards the supervised signal of purchases when representing reviews with the projected average word embeddings. 
%Moreover, since each review has its individual embedding, a huge number of parameters on reviews will be introduced to the model, which may lead to overfitting. 
Our experiments also show that projected average embeddings yield better results than paragraph vectors, so we exclude paragraph vectors from this paper. 

Another choice to represent reviews is using pre-trained BERT \cite{devlin2018bert} to encode the word sequence with transformers directly. However, we focus on modeling the interaction between the basic units, i.e., reviews, for both users and items, rather than capturing the semantic meaning carried in each review. Since reviews can be long and noisy, it is not for sure better to capture the reviews' complex semantic structures than to identify some keywords in the reviews using a reasonable and straightforward way. Also, encoding every review with BERT will introduce tremendous computation costs, which prevents us from using it. 

%where the relation that some reviews share a portion of the same words may not be sufficiently captured.
\subsection{Model Optimization}
Similar to previous studies \cite{ai2017learning,ai2019zero}, we optimize our model by maximizing the log likelihood of the observed (query,user,purchased item) triples, which can be written as:
\begin{equation}
\label{eq:optimization}
\begin{aligned}
\mathcal{L} &= \sum_{(q,u,i)} \mathcal{L}(q,u,i) = \sum_{(q,u,i)} \big ( \log P(i|q,u) + \log P(q,u) \big) \\
& \approx  \sum_{(q,u,i)} \log \frac{\exp(f(q,u,i))}{\sum_{i' \in S_i} \exp(f(q,u,i'))}
\end{aligned}
\end{equation}
where $f(q,u,\cdot)$ is computed with Eq. \ref{eq:item_score}, and $\log P(q,u)$ is ignored because it is predefined as a uniform distribution. 
%It is not feasible to optimize $\mathcal{L}$ directly due to the large number of items in the candidate set $S_i$. 
Due to the large number of items in the candidate set $S_i$, we adopt the negative sampling strategy \cite{le2014distributed,mikolov2013distributed} to estimate Eq. \ref{eq:optimization} 
and randomly select $k_{neg}$ negative samples from $S_i$ according to a uniform distribution. In addition, different L2 regularization settings could not improve the performance in our experiments, which indicates that overfitting is not a problem for our experiments. Hence, we do not include the regularization terms in Eq. \ref{eq:optimization}. 

\subsection{Model Interpretation}
\label{subsec:model_interpretation}
% Existing product search models \cite{ai2017learning, ai2019zero,guo2019attentive, bi2020transformer} usually explicitly represent user and item with embedding vectors and score a candidate item based on the matching between the item and the purchase intent built on the query and user profile. 
%Existing product search models either ignore fine-grained user-item interactions \cite{ai2017learning} or capture the interactions only at the item level \cite{ai2019zero, guo2019attentive, bi2020transformer} when building user profiles. 

Existing product search models \cite{ai2017learning,ai2019zero,guo2018multi,bi2020transformer} consider reviews associated with a user or item as a whole and do not differentiate their influences when matching users and items. In contrast, by conducting the attention mechanism on each unit in the sequence of the query, user reviews, and item reviews, \modelname{} can explicitly capture the interactions between the finer-grained units, i.e., query and user/item reviews. These fine-grained interactions reflect several essential aspects of the model:

\textbf{Explainable Item Matching.}
The attention scores of each item review concerning the query indicate which review plays a crucial role in matching this item with the purchase intent behind the query-user pair. We can rank the reviews with their attention weights from large to small for each retrieved item and display the ranking to the user. In contrast to showing item reviews according to their recency as most e-commerce platforms do, a system based on \modelname{} could help users understand why an item is retrieved by showing the most potentially helpful reviews, which may further facilitate their purchase decisions. 

\textbf{Dynamic Review Representation.}
\modelname{} can offer more potent learning abilities with multiple transformer layers, which could be beneficial when more interactions between reviews are needed. 
In a multiple-layer \modelname{}, the review embeddings associated with the user or the item are dynamically changed based on their interactions with the other units in the sequence $(q,R_u,R_i)$. In this case, more interactions happen between units when attending to $(q,R_u,R_i)$ with $r_{ui_{uk}} \in R_u $ and $r_{u_{ik'}i} \in R_i$. 

The final representation of a user review $r_{ui_{uk}}$ is learned from its interaction with $R_u$ and $R_i$. On the one hand, the self-attention mechanism of attending to $R_u$ with $r_{ui_{uk}}$ updates the embedding of $r_{ui_{uk}}$ by interacting with all the reviews from user $u$, where similarities and dissimilarities between reviews are taken into account. On the other hand, attending to $R_i$ with $r_{ui_{uk}}$ indicates that how the specific preference $u$ shows for $i_{uk}$ is satisfied by the descriptions of $i$ from other users' perspectives. Reviews in $R_i$ specify the other users' opinion towards item $i$, which could reflect $i$'s advantages and disadvantages user $u$ may care about. 
%The final representation of a user review $r_{ui_{uk}}$ is learned from its interaction with $R_u$ and $R_i$. On the one hand, attending to $R_u$ with $r_{ui_{uk}}$ takes similarities and dissimilarities between $r_{ui_{uk}}$ and each review in $R_u$ into account. On the other hand, each user review reflects the advantages and disadvantages of the item the user cares about. Attending to $R_i$ with $r_{ui_{uk}}$ indicates that how the specific preference $u$ shows for $i_{uk}$ is satisfied by the descriptions of $i$ from other users' perspectives. % shorten version

Similarly, the final vector of an item review $r_{u_{ik'}i}$ is also dynamically changed according to its interaction with $R_u$ and $R_i$. The interactions between $r_{u_{ik'}i}$ and $R_i$ readjust the representation of $r_{u_{ik'}i}$ by considering its relation with the other reviews in $R_i$. In addition, attending to the reviews in $R_u$ with $r_{u_{ik'}i}$ indicates how $r_{u_{ik'}i}$ matches the preferences expressed in the user's each historical review. This information and how each user preference is satisfied by item reviews carried in $r_{ui_{uk}}$ could be both beneficial to score item $i$.

\textbf{Dynamic Personalization}. 
%[In related work]%Previous personalized product search models either conduct personalization all the time \cite{guo2019attentive, ai2019explainable} or use a fixed coefficient to balance the effect of the query and personalization \cite{ai2017learning}. One notable exception is the zero attention model \cite{ai2019zero} that can control the influence of personalization dynamically by introducing a zero vector that the query can attend to when building a user profile. 
%In contrast to the zero attention model (ZAM) \cite{ai2019zero} where personalization can have at most equal effect as queries, 
%Similar to the model proposed in \cite{bi2020transformer}, \modelname{} can also control the influence of personalization from none to domination. 
% RTM has the "flexibility" or "capacity" to make predictions with variable emphasis on referencing the user's reviews
\modelname{} has the flexibility to make predictions with variable (none to full) emphasis on personalization, similar to \cite{bi2020transformer}, by learning different accumulative weights for the user's reviews. 
In \modelname{}, $q^{(l)}$ in Eq. \ref{eq:item_score} is computed from attending to the sequence of $(q,R_u,R_i)$ in the $(l-1)$-th layer with $q$. Personalization can take full effect when the attention weight assigned to $q$ is 0 and no effect when the reviews in $R_u$ have 0 accumulative attention weights. 

In addition, in contrast to \cite{ai2019zero,bi2020transformer} where personalization degree for a user depends only on her query, \modelname{} is more flexible since it could perform various degrees of personalization for different items given the same query-user pair since attention weights on $R_u$ could vary when $R_i$ is different even with the same $q$. This strategy allows the user profile to have differentiated effects when it contains much or no useful information while matching various items. 

\begin{comment}
\textbf{Differences from Other Transformer-based Product Search Models}. To the best of our knowledge, the only existing model that also uses transformers for product search is TEM \cite{bi2020transformer}. It encodes the sequence of a user and his/her purchased items with transformers to represent user search intents and learn the item embeddings by predicting words in the item reviews. However, there are significant differences between \modelname{} and TEM. First, TEM still follows the typical paradigm of existing product search models representing user intents and items explicitly with embedding vectors and conducting user-item level matching. Second, TEM models user intents with the user's purchased items and their corresponding representations built from all the reviews of each item. In \modelname{}, we also capture user intents from their purchase items, but we only use the current user's reviews for each purchased item to build the user intent representation.
%whose representations are learned based on their reviews from all users, while \modelname{} captures user intents with the reviews from the current user and build dynamic item representations accordingly. 
Third, while TEM constructs the user vectors dynamically, the item vectors are static all along.
In contrast, \modelname{} construct both user vectors and item vectors dynamically based on search context.
% Also, \modelname{} has segment embeddings to differentiate user and item reviews which do not exist in TEM. 
\end{comment}

	%!TEX root=main.tex
\section{Experimental Setup}
\label{sec:exp_setup}
In this section, we first show how we construct the datasets and conduct evaluation, then we describe the baseline models and training settings for different models. 
\subsection{Datasets and Evaluation}
\label{subsec:dataset_eval}
The Amazon product search dataset \cite{mcauley2015inferring} \footnote{\url{http://jmcauley.ucsd.edu/data/amazon/}} is the only available dataset for product search that have user reviews. % although queries need to be synthesized. 
As in previous work \cite{ai2017learning, guo2019attentive, van2016learning,zhang2018towards, ai2019explainable, bi2019conversational, bi2020transformer}, we use it for our experiments. Specifically, we use the 5-core data \cite{mcauley2015inferring} where each user and each item has at least 5 associated reviews. Our experiments are based on three categories of different scales, which are \textit{Clothing, Shoes \& Jewelry}, \textit{Sports \& Outdoors}, and \textit{CDs \& Vinyl}. The statistics are shown in Table \ref{tab:stats}. 

\textbf{Query Construction}. 
Following the same paradigm used in \cite{ai2017learning, guo2019attentive, ai2019explainable, van2016learning,zhang2018towards, bi2019conversational, bi2020transformer}, we construct queries for each purchased item with the product category information. 
This strategy is based on the finding that directed product search is users' search for a producer's name, a brand, or a set of terms describing product category \cite{rowley2000product}. Precisely, we extract the multi-level category information from the meta-data, concatenate the words in the categories, and remove stopwords and duplicate words to form a query string. 
Since an item could belong to multiple categories, there may be multiple extracted queries for the item. Each query is considered as the initial query issued by the user and leading to purchasing the item. The queries are general and do not reveal specific details of the purchase items. 

\begin{table}
	\caption{Statistics of the Amazon datasets.}
	\centering
	\label{tab:stats}  
	\small
	%   \addtolength{\tabcolsep}{3pt} 
	\scalebox{0.95}{    
		\begin{tabular}{p{0.8cm}  r  r  r }
			\hline
			%Dataset & Sports &  Clothing & CD \\
			%Dataset & Sports \& Outdoors & Clothing, Shoes \& Jewelry & CDs \& Vinyl \\
			\multirow{2}{*}{Dataset} & Sports \&  & Clothing, Shoes \&  & CDs \&  \\
			&  Outdoors &  Jewelry &  Vinyl \\
			\hline
			\#Users & 35,598 & 39,387 & 75,258 \\
			\#Items & 18,357 & 23,033 & 64,443 \\
			\#Reivews & 296,337 & 278,677 & 1,097,591 \\
			\#Queries & 1,538 & 2,021 & 695 \\
			\#Vocab & 32,386 & 21,366 & 202,959 \\
			ReviewLen & 89.18$\pm$106.99 & 62.22$\pm$60.16 & 174.56 $\pm$177.05 \\
			QueryLen & 7.07 $\pm$ 1.74 & 7.14$\pm$1.97 & 5.77$\pm$1.65 \\
			%\multicolumn{4}{l}{Training/Validation/Test Splits} \\
			%\#q-u pairs & 269,850/776/910 & 467,651/4,106/4,025 & 1,524,168/10,930/10,077\\
			\hline
			\multicolumn{4}{l}{\#Query-user pairs} \\
			Train & 269,850 & 467,651 & 1,524,168 \\ 
			Valid/Test & 776/910 & 4,106/4,025 & 10,930/10,077 \\
			\hline
			\multicolumn{4}{l}{\#Purchased items per query-user pair} \\
			Train & 1.16$\pm$0.55 & 1.30$\pm$0.82 & 2.95$\pm$8.53 \\
			Valid/Test & 1.01$\pm$0.08/1.02$\pm$0.24 & 1.00$\pm$0.08/1.00$\pm$0.05 & 1.12$\pm$0.61/1.11$\pm$0.56 \\
			\hline
		\end{tabular}
	}
\end{table}
\textbf{Training/Validation/Test Splits}. 
As in \cite{bi2020transformer}, we split each dataset into training, validation, and test sets according to the following steps. First, we randomly put 70\% queries in the training set, and the rest 30\% are shared by the validation and test sets so that none of the test queries have been seen during training. 
%To construct our experimental testbed, we split each dataset into training, validation and test sets according to the following steps. First, we randomly divide all the available queries according to the ratio 0.7:0.3. 70\% queries are put in the training set and the rest 30\% are shared by the validation and test sets. 
%For any purchased item, if all the associated queries fall into the test set, we randomly select one query and put it back to the training set. In this way, we ensure that none of the test queries have been seen during training. 
Then, we partition each user's' purchases to training, validation, and test set according to the ratio 0.8:0.1:0.1 in chronological order. 
% For an arbitrary purchase in the validation and test set, 
If none of the queries associated with the purchased item is in the test set, the purchase will be moved back to the training set. 
In contrast to randomly partitioning data into training and test set as in previous work \cite{ai2017learning,bi2019conversational,ai2019explainable,zhang2018towards}, our dataset is closer to a real scenario, where all the purchases in the test set happen after the purchases in the training set. This also makes retrieval on our test set harder since future information can be used to predict past purchases in the previous datasets. Also, our training and test sets have less similar distributions compared with previous datasets, which makes model prediction more difficult as well. 

%\textbf{Evaluation Methodology.}
%For fair comparisons, we evaluate the ranking lists of the entire item set in each category ranked by both our \modelname{} and all the baselines. Although \modelname{} is designed for re-ranking top results, we have to run it on all the items due to two reasons: (1) If we want to limit the number of candidate items without using a retriever, we must use sampling strategies to manually create candidate sets with positive items and sampled negative items, which has been shown to be not trustworthy in terms of comparing different retrieval models~\cite{krichene2020sampled}; (2) Even with the best retrieval model today, the top 100 or 1000 candidate items retrieved by the retriever may include none positive item for many queries in the Amazon product search dataset due to data sparsity, which would limit the performance upper bound of baselines and our models in the experiments. 
%Therefore, despite the fact that ranking all items in the Amazon product search dataset would result in slow inference steps and low metric scores, we still run all baselines and our models in this full ranking setting to compare their effectiveness for product search.
\textbf{Evaluation Metrics.}
Following the typical way of collecting candidates with an efficient initial ranker and using neural models for re-ranking in document retrieval, we re-rank the candidate items retrieved by BM25 \cite{robertson1995okapi} with each method and obtain the ranking lists. \footnote{In the experiments where each method ranks all the items in the collections, we have similar observations, so we do not include this setting for space concerns. } Then we use Mean Reciprocal Rank (MRR), Normalized Discounted Cumulative Gain at 20 (NDCG@20), and Recall at 20 (R@20) as evaluation metrics. MRR shows the first position where any purchased item is retrieved; NDCG@20 focuses on the top 20 items' ranking performance where higher positions have more credits, and R@20 indicates how many target items are retrieved in the top 20 results in total. Fisher random test \cite{smucker2007comparison} with $p < 0.05$ is used to measure significant differences between results. 

\subsection{Baselines}
We compare our \modelname{} with eight representative baselines:

	\textbf{BM25}: The BM25 \cite{robertson1995okapi} model is based on word matching between queries and item reviews, which also provides the initial ranking lists for the other models. 
	
	\textbf{POP}: The Popularity (POP) model ranks items according to their frequency of being purchased in the training set. 
	
	\textbf{LSE}: The Latent Semantic Entity model (LSE) \cite{van2016learning} is an embedding-based non-personalized model that learns the vectors of words and products by predicting the products with n-grams in their reviews. It then scores the products with the cosine similarity between their vectors and query vectors. 
	
	\textbf{QEM}: The Query Embedding Model (QEM) \cite{ai2019zero}, is also a non-personalized model that conducts item generation based on the query embedding, and item embeddings are learned by predicting words in their associated reviews. 
	
	\textbf{HEM}: The Hierarchical Embedding Model (HEM) \cite{ai2017learning} has the item generation model and language models of users and items. It balances the effect of personalization by applying a convex combination of user and query representation. 
	
	\textbf{AEM}: The Attention-based Embedding Model (AEM) \cite{ai2019zero,guo2019attentive} constructs query-dependent user embeddings by attending to users' historical purchased items with the query. \footnote{The attention models described in \cite{ai2019zero} and \cite{guo2019attentive} are highly similar to each other, so we only implement the one in \cite{ai2019zero} and named it as AEM. }
	%The attention model proposed by \citet{guo2019attentive} is similar to this model. 

	\textbf{ZAM}: The Zero Attention Model (ZAM) \cite{ai2019zero} extends AEM with a zero vector and conducts differentiated personalization by allowing the query to attend to the zero vector.
	
	\textbf{TEM}: The Transformer-based Embedding Model (TEM) \cite{bi2020transformer} is a state-of-the-art model that encodes query and historically purchased items with a transformer and does item generation based on the encoded query-user information. 
	% Item embeddings are also learned from predicting words in the item reviews. We compare \modelname{} with TEM to see whether it is better to match the query-user pair and item at the review level compared with the item level based on the transformer architecture.     
%TODO: add more baselines QL and POP. 

BM25, POP, LSE, and QEM are non-personalized retrieval models, and all the rest are personalized product search models. 
%We use the top 100 results from BM25 for re-ranking on $Sports$ and $Clothing$ whose recall values are 0.425 and 0.343, respectively. On $CDs$, the $Recall@100$ of BM25 is only 0.108, so we use the top 1000 items for re-ranking, which has a higher recall - 0.370.
%Besides POP, we only include neural models as our baselines since exact term matching has been shown to be not enough to retrieve ideal products and term-based retrieval models are much less effective in previous studies on product search \cite{ai2017learning, bi2019conversational, ai2019zero}. 

\subsection{Implementation Details}
\label{subsec:train_setting}
%We trained both our model and all the neural baseline models on a single NVIDIA TITAN X GPU with 12GB memory. 
All the baselines were trained for 20 epochs with 384 samples in each batch according to the settings in their original papers \cite{ai2017learning, ai2019zero, bi2020transformer}, and they can converge well. We trained our model for 30 epochs with 128 samples in each batch. In the baseline models, each word in the reviews of a target item corresponds to one entry (word, item, user, query) in the batch, while in \modelname{} each target item has one entry (item, user, query) in a batch.
The number of negative samples for items or words in all the models is set to 5. We set the embedding size of all the models to 128. Larger embedding sizes do not lead to significant differences, so we only report results with $d=128$.  The sub-sampling rate of words in all the neural baseline models is set to $1e-5$. We sweep the number of attention heads $h$ from \{1,2,4,8\} for AEM, ZAM, TEM, and our \modelname{}. For TEM and \modelname{}, we vary the number of transformer layers $l$  from 1 to 3 and set the dimension size of the feed-forward sub-layer of the transformer from \{128, 256, 512\}. We cutoff reviews to 100 words and limit the number of historical reviews for a user and an item, i.e., $m$ and $n$ in Figure \ref{fig:rtm}, to 10 and 30, respectively. We use Adam with $\beta_1=0.9, \beta_2=0.999$ to optimize \modelname{}. The learning rate is initially set from \{0.002, 0.005, 0.01, 0.02\} and then warm-up over the first 8,000 steps, following the paradigm in \cite{vaswani2017attention, devlin2018bert}. To make the training of \modelname{} more stable, we initialize the parameters of words with embeddings pre-trained with Word2Vec \cite{mikolov2013distributed}. For the number of candidates, we use the top 100 results from BM25 for re-ranking on $Sports$ and $Clothing$ whose recall values are 0.425 and 0.343, respectively. On $CDs$, the $Recall@100$ of BM25 is only 0.108, so we use the top 1000 items for re-ranking, which has a higher recall - 0.370. Our code can be found at \url{https://github.com/kepingbi/ProdSearch}.
%We will release our code after the paper is published. 

	%!TEX root=main.tex
\section{Results and Discussion}
%TODO: add a starting paragraph
\label{sec:results}
%TODO{Revise This}
In this section, we first compare the overall performance of \modelname{} and the baseline models. Then we conduct model analysis and case studies to interpret the model behavior. 

%Then we show model analysis on several aspects. At last, we conduct case studies to interpret the model behavior. 
\begin{table*}
	\caption{Comparison between the baselines and our proposed \modelname{}. `*' marks the best baseline performance. `$\dagger$' indicates significant improvements over all the baselines in Fisher Random test \cite{smucker2007comparison} with $p<0.05$. }%, The best performance is highlighted in boldface. } 
	\label{tab:overallperf}
	\small
	%\large
	%    \addtolength{\tabcolsep}{3pt} 
	%    \scalebox{0.94}{    
	\begin{tabular}{ c | l || l | l | l || l | l | l || l | l | l   }
		%        \begin{tabular}{  p{1.5cm} || p{1.5cm}  || p{1.5cm}  ||  p{1.5cm}  || p{1.5cm}  | p{1.5cm}  | p{1.5cm}  || l || l ||  l | l | l  }
		\hline
		\multicolumn{2}{c||}{Dataset}& \multicolumn{3}{c||}{Sports \&  Outdoors} & \multicolumn{3}{c||}{Clothing, Shoes \&  Jewelry} & \multicolumn{3}{c}{CDs \& Vinyl} \\
		\hline
		\multicolumn{2}{c||}{Model} & $MRR$ & $NDCG@20$ & $R@20$ &  $MRR$ & $NDCG@20$ & $R@20$ & $MRR$ & $NDCG@20$ & $R@20$\\
		\hline
		\multirow{4}{*}{Non-personalized} 
        %& QL &  0.040 & 0.060 & 0.008 & 0.038 & 0.056 & 0.007 & 0.009
        & BM25 & 0.049 & 0.051 & 0.173 & 0.044 & 0.051 & 0.160 & 0.011 & 0.015 & 0.042 \\ 
        & POP & 0.033 & 0.055 & 0.158 & 0.030 & 0.044 & 0.112 & 0.015 & 0.018 & 0.039 \\ 
        & LSE & 0.026 & 0.047 & 0.148 & 0.041 & 0.058 & 0.135 & 0.017 & 0.021 & 0.044 \\ 
        & QEM & 0.049 & 0.070 & 0.172 & 0.039 & 0.057 & 0.144 & 0.010 & 0.014 & 0.037 \\ 
        \hline
        \multirow{8}{*}{Personalized}
		& HEM & 0.044 & 0.071 & 0.197 & 0.043 & 0.061 & 0.144 & 0.021 & 0.030 & 0.073 \\ 
        & AEM & 0.047 & 0.076 & 0.209 & 0.048 & 0.069 & 0.162 & 0.023 & 0.033 & 0.084 \\ 
        & ZAM & 0.052 & 0.083 & 0.220 & 0.047 & 0.069 & 0.166 & 0.025 & 0.036 & 0.086 \\ 
        & TEM & 0.060* & 0.094* & 0.238* & 0.052* & 0.076* & 0.182* & 0.026* & 0.038* & \textbf{0.095}* \\ 
        \cline{2-11}
		& \modelname{} ($\mathcal{I}_{pos}\!\!=\!\!0, \!\mathcal{I}_{seg}\!\!=\!\!0$)
		& 0.065 & 0.092 & 0.208 & 0.061$^\dagger$ & 0.087$^\dagger$ & 0.190$^\dagger$ & 0.027 & 0.036 & 0.084 \\ 
		& \modelname{} ($\mathcal{I}_{pos}\!\!=\!\!0,\! \mathcal{I}_{seg}\!\!=\!\!1$)
		& 0.058 & 0.093 & \textbf{0.242} & 0.068$^\dagger$ & 0.099$^\dagger$ & \textbf{0.224}$^\dagger$ & \textbf{0.030}$^\dagger$ & \textbf{0.042}$^\dagger$ & \textbf{0.095} \\ 
		& \modelname{} ($\mathcal{I}_{pos}\!\!=\!\!1, \!\mathcal{I}_{seg}\!\!=\!\!0$)
		& 0.082$^\dagger$ & 0.110$^\dagger$ & 0.234 & \textbf{0.071}$^\dagger$ & \textbf{0.101}$^\dagger$ & 0.219$^\dagger$ & \textbf{0.030}$^\dagger$ & 0.039 & 0.085 \\ 
		& \modelname{} ($\mathcal{I}_{pos}\!\!=\!\!1 , \!\mathcal{I}_{seg}\!\!=\!\!1$)
        & \textbf{0.096}$^\dagger$ & \textbf{0.123}$^\dagger$ & 0.237 & 0.069$^\dagger$ & 0.099$^\dagger$ & 0.218$^\dagger$ & 0.028 & 0.040 & 0.094 \\
        \hline
	\end{tabular}
	%}
\end{table*}

\begin{figure*}
	\centering
	\begin{subfigure}{.33\textwidth}
		\includegraphics[width=2.35in]{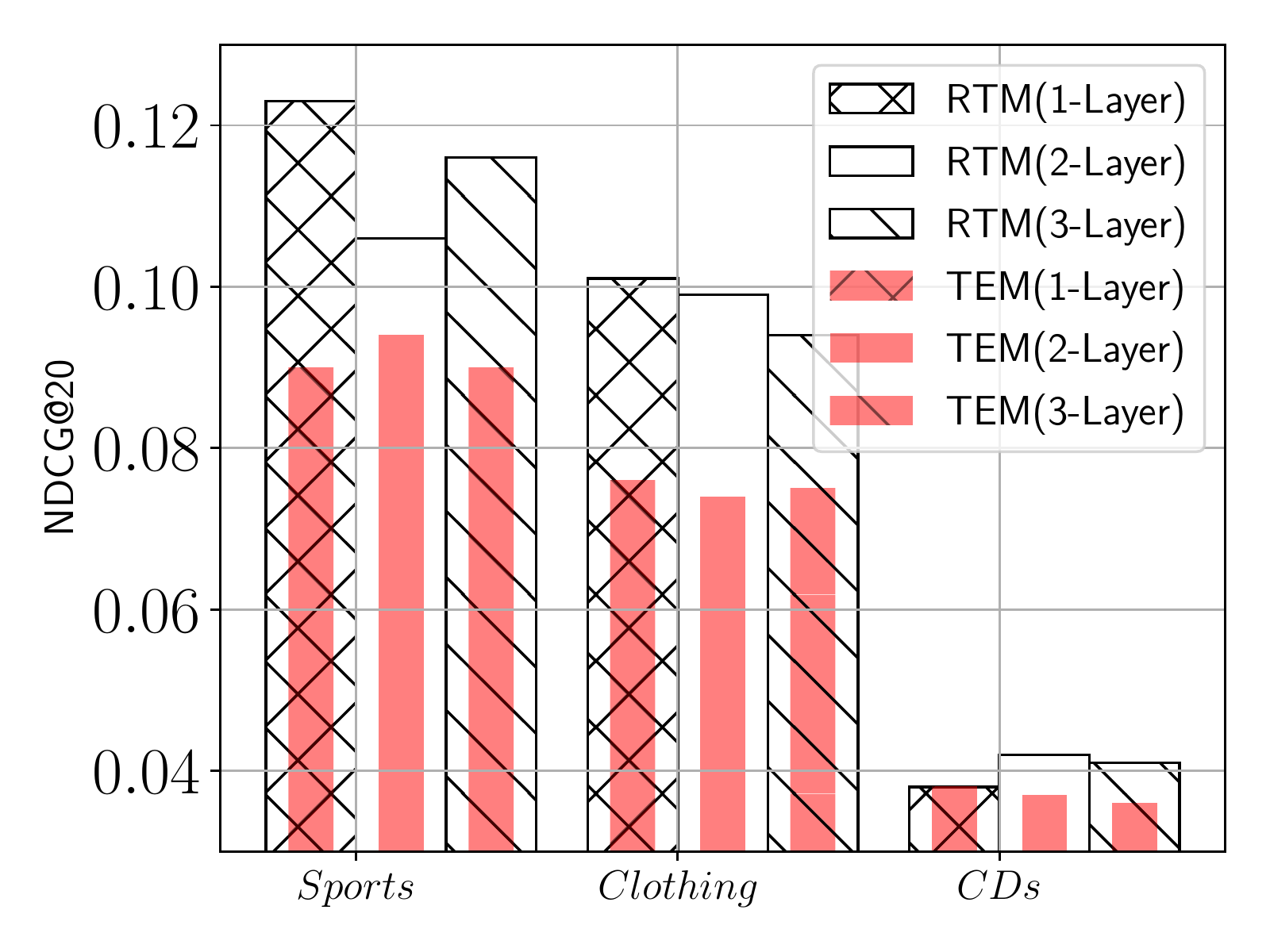}
		\caption{Effect of Transformer Layer Numbers}
		\label{fig:transformer_layers}
	\end{subfigure}%
	\begin{subfigure}{.33\textwidth}
		\includegraphics[width=2.35in]{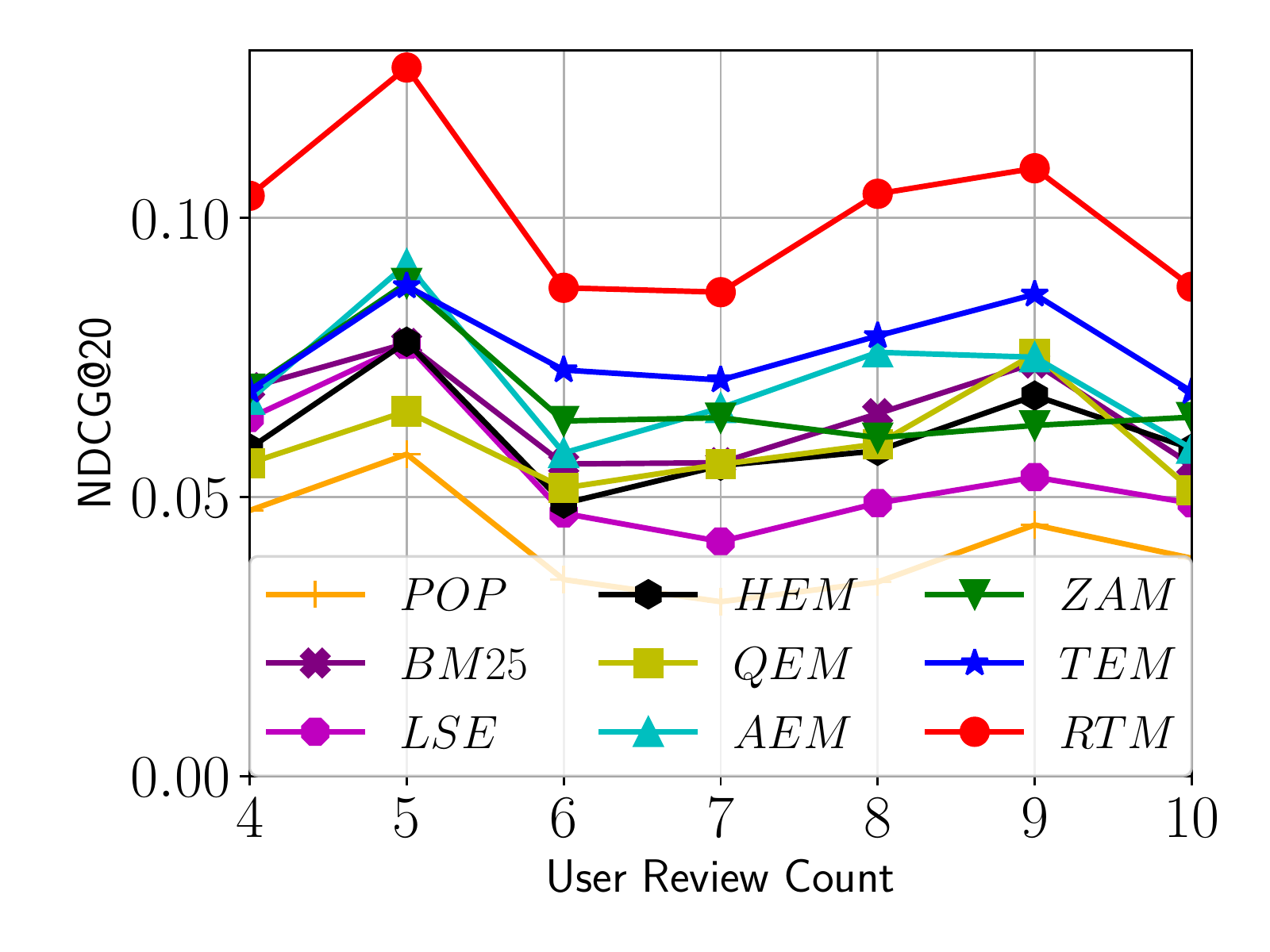}
		%\caption{The MRR of different feedback models with respect to feedback iterations on \textit{Cell Phones\&Accessories}.}
		\caption{Effect of User Review Counts}
		\label{fig:clothing_perf_distr}
	\end{subfigure}%
	\begin{subfigure}{.33\textwidth}
		\includegraphics[width=2.35in]{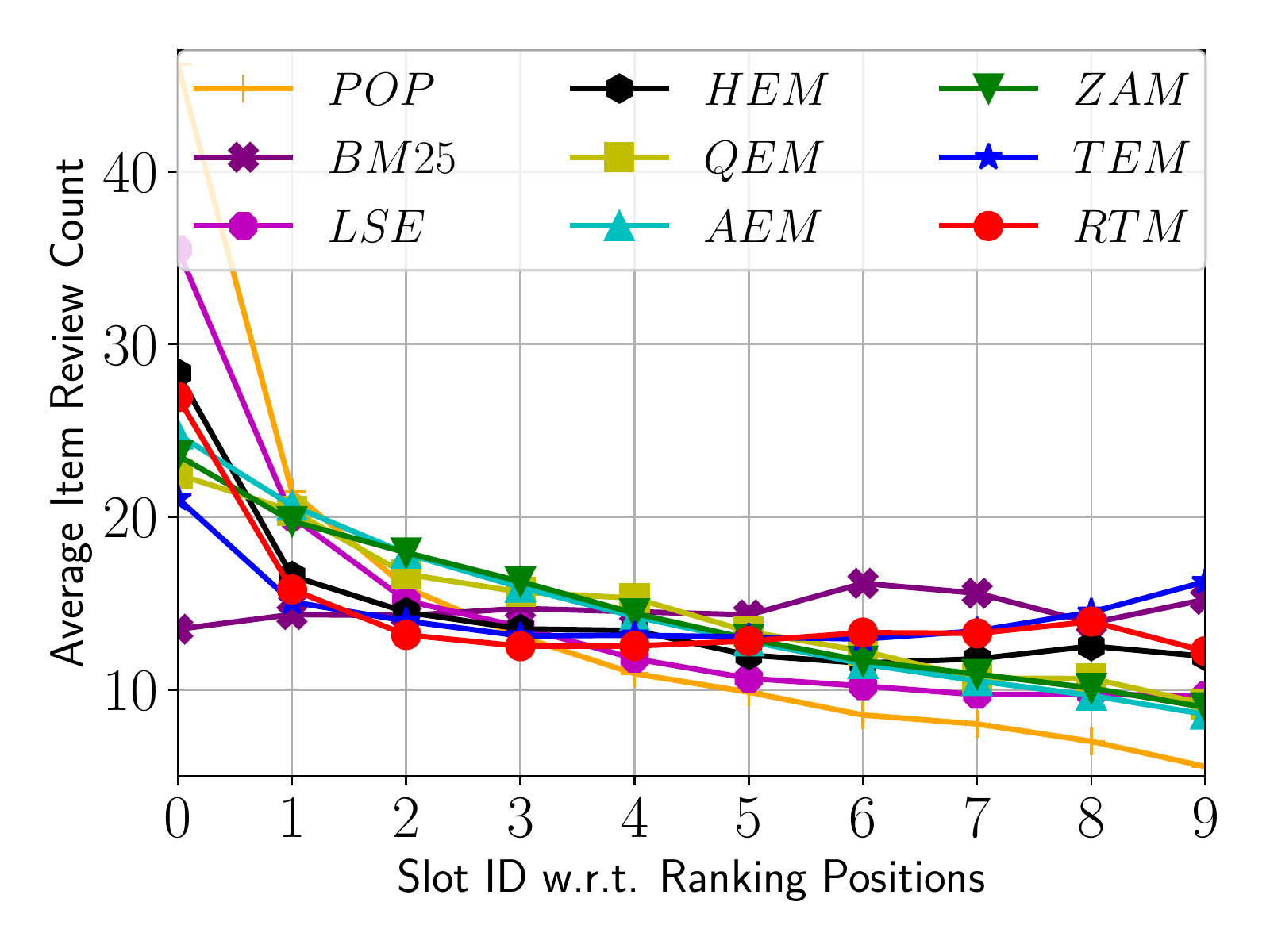}
		\caption{Item Review Count w.r.t. Ranking Positions}
		\label{fig:clothing_rcount_distr}
	\end{subfigure}%
	\caption{Model analysis in terms of different aspects. Figure \ref{fig:clothing_perf_distr} and \ref{fig:clothing_rcount_distr} correspond to Clothing, Shoes $\&$ Jewelry. }
\end{figure*}

\subsection{Overall Performance}
\label{subsec:overall_perf}
Table \ref{tab:overallperf} shows each system's overall ranking performance on the three datasets. We show four variants of \modelname{} that use both, either, or none of the position embeddings and segment embeddings by setting $\mathcal{I}_{pos}$ and $\mathcal{I}_{seg}$ in Eq. \ref{eq:q0} to 0 or 1. We illustrate the effect of position and segment embeddings in Section \ref{subsec:model_analysis}. Note that the numbers in Table \ref{tab:overallperf} are small for two reasons: 1) there is only about 1 target purchased item out of 20k\textasciitilde65k items for each query-user pair in the test sets, shown in Table \ref{tab:stats}. % Thus there are no positive items in the candidate sets constructed with BM25 for about 60\%-70\% queries on each dataset. 
% Note that the numbers in Table \ref{tab:overallperf} are small since there is only about 1 target purchased item out of 20k\textasciitilde65k items for each query-user pair in test sets, shown in Table \ref{tab:stats}. 
2) as we mentioned in Section \ref{subsec:dataset_eval}, the search task on our sequentially split data is more challenging than on the randomly divided partitions in \cite{ai2017learning, bi2019conversational, zhang2018towards}.
% which is another reason for the small performance numbers. 

As in previous studies \cite{ai2017learning, ai2019zero, guo2019attentive}, we observe that non-personalized models perform worse than personalized models. BM25 performs better on $Sports$ and $Clothing$ than $CDs$, which indicates that term matching plays a more important on $Sports$ and $Clothing$. In contrast, POP matters more on $CDs$ than $Sports$ and $Clothing$. Non-personalized neural models do not always outperform BM25, especially on $Sports$ and $Clothing$, probably because semantic matching brings limited benefits when most candidate results from BM25 have exact term matching. 
%Also, the popularity information carried in the embeddings has limited contribution to the first two datasets as shown by POP. 

%As in previous studies \cite{ai2017learning, ai2019zero, guo2019attentive}, we observe that non-personalized models perform worse than personalized models. POP hardly retrieves any target results in all the categories, which shows that the group preference cannot satisfy individual needs. QEM is better than LSE on Sports \& Outdoors and similar on the other two datasets. Both of them outperform POP by a large margin by considering the product relevance to test queries.

Among the personalized retrieval baselines, TEM achieves the best performance, which is consistent with \cite{bi2020transformer} and confirms the benefit of using transformers.
ZAM performs better than AEM most of the time, indicating that dynamic personalization is helpful. Similar to \cite{ai2019zero, guo2019attentive}, we also observe that HEM could not outperform the attention-based models, which indicates that building a dynamic user profile helps improve the result quality compared with using static user representation across the search sessions. 

On all the categories, \modelname{} achieves the best performance. It has significant improvements upon the best baseline -- TEM -- in most metrics on all the datasets. 
%On $Sports$ and $CDs$, with similar numbers of positive items ranked to top 20, \modelname{} promotes them to higher positions. On $Clothing$, \modelname{} also identifies more target items in the top 20 results. 
It confirms that by modeling the dynamic matching between user and item, \modelname{} has significant advantages over models with static item profiles, although some of them also have dynamic user profiles (AEM, ZEM, and TEM). In addition, by capturing the finer-grained matching at the review level, \modelname{} can differentiate different items better than matching them at the item level in TEM. 

% interactions between a query-user pair and an item in finer-grained units, i.e., reviews,
%On all the categories, \modelname{} achieves the best performance in terms of all the three metrics. Our best model outperforms all the baselines significantly and the improvement over the best baseline on each dataset is more than 20\%. The observation that MRR, NDCG, and Precision are all improved indicates that \modelname{} can identify more target items in the top 20 results and promote them to higher positions. It also confirms that by modeling the dynamic matching between user and item, \modelname{} has significant advantages over models that have static item profiles although some of them also have dynamic user profiles (AEM, ZEM, and TEM). In addition, by capturing interactions between a query-user pair and an item in finer-grained units, i.e., reviews, \modelname{} can differentiate different items better than matching them at the item level in TEM. 

%\subsection{Effect of Different Transformer Settings}
\subsection{Model Analysis}
\textbf{Position and Segment Embeddings.}
\label{subsec:model_analysis}
As shown in Table \ref{tab:overallperf}, \modelname{} without position and segment embeddings has the worse performance most of the time, showing the necessity to differentiate the reviews from the user and the item. On $Sports$ and $Clothing$, position embeddings are always helpful, 
which indicates that the user's recent purchases have different influences from the long-ago purchases. The latest reviews reveal more accurate information about the products. For example, users' preferences on clothes styles may change according to the seasonal trend; a swimming earplug product has been updated lately, and recent customer reviews complained the new version is not as comfortable as before. In these cases, position embeddings that capture reviews' recency can help identify the current user preference and item status and potentially improve the search quality. 

In contrast, on $CDs$, incorporating position embeddings does not lead to better evaluation results than using segment embeddings alone. This shows that the order of user and item reviews do not need to be differentiated as long as we know which of them corresponds to the user and item respectively. This observation is consistent with our intuition that the content in a CD and sound quality are usually static regardless of the review order. We can also infer that long-ago purchases play similar roles as recent purchases in terms of representing user preferences on $CDs$.

Using both position and segment embeddings does not always lead to the best evaluation results. The possible reasons are that the low and high positions can indicate which sections of the input correspond to the user and item respectively, which makes segment embeddings not necessary sometimes. When the chronological order of reviews is not crucial in some categories, the sequence of reviews does not matter so that position embeddings could introduce noise to the model and harm the performance, as we mentioned in Section \ref{subsec:personal_score}. 

\textbf{Number of Layers.} \modelname{} achieves the best performance with 2 layers on $CDs$ and 1 layer on $Sports$ and $Clothing$. This indicates that the dynamic review representation introduced in Section \ref{subsec:model_interpretation} is beneficial on larger datasets. As shown in Table \ref{tab:stats}, $CDs$ has more average reviews per user/item than the other two datasets, and it also has a larger vocabulary, which makes the contexts of each review more varied. We will further analyze the behavior of \modelname{} with single and multiple layers in Section \ref{subsec:case_study} by case studies. 

%Another observation is that TEM and \modelname{} needs different numbers of layers to achieve the best performance on the same dataset. When we encode information at various levels, i.e., the sequence of (query, users' purchased items) at the item level or the sequence of (query, user reviews, item reviews) at the review level, different amounts of interactions are required.

\begin{table*}
	\caption{A case of query, user, and purchased item for the single-layer RTM case study. The attention weights are average from all the 8 attention heads. }
	\centering
	\label{tab:clothing_case} 
	%\scriptsize
	\footnotesize
	%   \addtolength{\tabcolsep}{3pt} 
	%    \scalebox{0.95}{    
	%\begin{tabulary}{\linewidth}{LCL|LCL}
	\begin{tabular}{C{0.4cm} | C{0.2cm} |p{6cm} | C{0.4cm} | C{0.2cm} | p{8.5cm} }
		\hline
		\multicolumn{6}{c}{Query: \textbf{"clothing shoe jewelry men big tall active athletic sock"} (Attention weight w.r.t. $q^{(0)}$:0.031)} \\
		\hline
		%            \multicolumn{3}{c|}{User Reviews} & \multicolumn{3}{c}{Item Reviews} \\
		%            \hline
		Attn & ID & User Reviews& Attn & ID & Item Reviews \\
		\hline
		0.027 & ur3 & Fruit of the Loom \textbf{T-Shirt}. This shirt is a great value for the price. \textbf{It is snug and fits me perfectly}. There is enough room to wear the shirt with an under-shirt as well, giving me warmth. &
		0.042 & ir9 & As expected bought these socks for my \textbf{husband} because he is always wearing holes in his socks, and I am looking to Carhartt to provide a sock that might be able to better withstand his abuses. So far they are \textbf{sufficiently cushiony}, \textbf{they stay up on his legs} and get the job done. Time will tell if they are as durable as I am hoping. \\
		\hline
		%            0.212 & ur3 & review &
		%            0.334 & ir9 & review \\
		%            0.179 & ur4 & Hanes Men's T-Shirt This shirt is a good value for the amount spent.  It is snug and warm and  I am happy that I purchased it. &
		%            0.326 & ir10 & Great Socks I bought a pair of these on the spur of the moment as a stocking stuffer 3 years ago.  Since then that is all my husband will wear.  He says these make his feet feel better.  They last longer than what he had been wearing.  Had to buy more this year because my sons wear them now too.  Carhartt brand is hard to beat. \\            
		%            0.162 & ur5 & Carhart Watch Hat This cap is a perfect fit, and well worth the price.  Why didn't I buy it previously and wear it? &
		%            0.316 & ir19 & best ever ! These are just the best socks!  They are a great weight and have stayed white after several washes.  I expect to buy more when these start to wear out. Recommend to any guy who wants a pair of quality socks. \\
		%            0.124 & ur1 & Feedback on moccassins Nice house shoes.  Soft, nice color and comfortable,  Very durable, and please to wear. Would purchase again when these are worn out. &
		%            0.156 & ir7 & Super good! The socks were as described, really good material and tight snugly fit. If you wash them with a good detergent they keep the nice smell for long, indicative of the fabric's quality. \\
		%            0.078 & ur2 & review & 
		%            0.143 & ir4 & review \\
		0.010 & ur2 & Casio Men's MQ24-1E black resin \textbf{watch}. I love it because it is small, easy to fasten, lightweight, and inexpensive. I had to cut the band for a smaller wrist, but glad I bought the watch. & 
		0.018 & ir4 & Can't go wrong with Carhartt. These are great socks. There are great for everyday work. They're comfortable and durable. It's a great product Not much more to say. \\
		
		\hline
	\end{tabular}
	%\end{tabulary}
	%    }
\end{table*}

\textbf{User Review Count.}
Figure \ref{fig:clothing_perf_distr} shows the performance of query-user pairs with different counts of user reviews in the training set of $Clothing$. The other two datasets show similar trends. The minimum number is 4 since each user has at least five reviews and 10\% of them has been put to the validation or test set, as mentioned in Section \ref{subsec:dataset_eval}. According to Section \ref{subsec:train_setting}, the max number is 10 since we use at most 10 historical reviews for users. The corresponding numbers of the query-user pairs with user review count from 4 to 10 are 1297, 655, 453, 417, 243, 180, and 780, respectively. We can see that \modelname{} has consistent improvements over other models on query-user pairs with different review counts and \modelname{} can achieve compelling performance with a decent small number of user reviews. 

\textbf{Item Review Count w.r.t. Ranking Positions.}
Figure \ref{fig:clothing_rcount_distr} compares different methods in terms of their tendency to rank items with more reviews to higher positions, i.e., their preferences on popularity. We only show $Clothing$ since the other two datasets have similar trends. For the top 100 items ranked by each method, we group them into 10 slots and show the items' average review counts in each slot. For example, Slots 0 and 1 correspond to items that are ranked from 1 to 10 and 11 to 20. We observe that BM25 and POP have the least and most tendency to value popular items respectively, which is consistent with their principles of only emphasizing relevance or popularity. Among other methods, LSE emphasizes popularity the most; HEM and \modelname{} put more popular items to top 10 and less popular items to positions from 11 to 50 than ZAM, AEM, and QEM; TEM ranks the fewest popular items to top 10 and has similar value to \modelname{} at Slot 1. Overall, the fact that \modelname{} achieves better performance than baselines suggests that \modelname{} could better balance popularity and relevance by the review-level matching. 

\subsection{Case Study}
\label{subsec:case_study}
We sample two cases in the test set of $Clothing$ and $CDs$ from our best model to illustrate the three aspects of \modelname{} introduced in Section \ref{subsec:model_interpretation} and show how \modelname{} identifies useful information from review-level user-item interactions. 
We show one example in $Clothing$ to represent the case of a single-layer \modelname{}, and the other example in $CDs$ to illustrate how a multi-layer \modelname{} performs.

%We show one example in $Clothing$ to represent the case when one transformer layer is the best for \modelname{}, and the other example in $CDs$ is to show how the model performs with multiple transformer layers. 

\begin{figure}[t]
	\centering
	\includegraphics[width=0.45\textwidth]{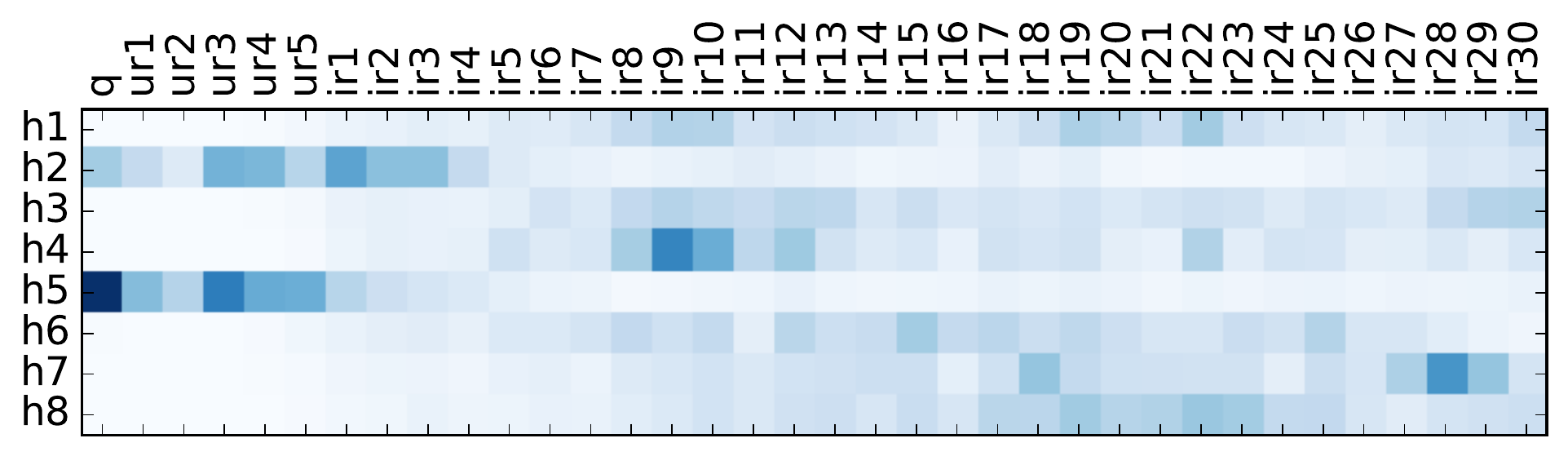} %
	\caption{Attention weights for the case shown in Table \ref{tab:clothing_case}.}
	\label{fig:clothing_attn}
\end{figure}

\textbf{Case Analysis for Single-layer \modelname{}}. 
%\textbf{Case in Clothing, Shoes \& Jewelry}. 
% We first show how \modelname{} allocates attention to the sequence of the query, user's historical reviews, and reviews of an ideal item for the query ``clothing shoe jewelry men big tall active athletic sock'' in Figure \ref{fig:clothing_attn}. Since the best model for $Clothing$ has only 1 transformer layer, all the attention weights shown in the figure are computed with respect to $q^{(0)}$ in Figure \ref{fig:rtm}. There are 8 attention heads in total and they capture different aspects of the sequence. As shown in Figure \ref{fig:clothing_attn}, $h2$ and $h5$ focus more on how to allocate attention to each user reviews, while the other heads concentrate on differentiating important item reviews. 
In Figure \ref{fig:clothing_attn}, we show how \modelname{} allocates attention to the sequence of the query, user's historical reviews, and reviews of an ideal item for the query ``clothing shoe jewelry men big tall active athletic sock''  with respect to $q^{(0)}$. There are 8 attention heads in total and they capture different aspects of the sequence. $h2$ and $h5$ focus more on how to allocate attention to each user reviews, while the other heads concentrate on differentiating important item reviews. 
Overall, the item reviews have the most portion of accumulative attention weights from the query, which is reasonable since item reviews are more important to differentiate candidate items. The accumulative attention weights on the user side are positive, which shows that user's historical reviews do help and personalization is needed. 
%The accumulative attention weights from all the user reviews indicate how much personalization can take effect given the query and the item. If no historical information from the user could help match the query-user pair with the item, there could be no personalized information used for the matching. In this case, some of the user's historical reviews do help, so the accumulative attention weights on the user side are positive.

%%In addition, although without segment embeddings on $Clothing$, \modelname{} can still identify whether the reviews are from the user or the item. As we mentioned in Section \ref{subsec:transformer_setting}, \modelname{} may learn the relationship from positional embeddings. 

%If we rank each unit in the sequence with their average attention weights from the 8 heads, the query itself is ranked at 15 among the 36 units, which indicates that the query has a medium effect on the final matching score. $ur3$ (the 3rd user review) and $ir9$ (the 9th item review) have the most attention scores among the user reviews and item reviews respectively. In contrast, $ur2$ and $ir4$ have the least attention weights. Note that these attention scores are dynamically changed given the reviews of a different user or item. 

To compare the reviews that take the most and least effect on matching the item with the query-user pair, we show the review text of $ur3$, $ur2$, $ir9$, and $ir4$ (the 3rd, 2nd user review and the 9th, 4th item review) and their attention weights in Table \ref{tab:clothing_case}. The query ``clothing shoe jewelry men big tall active athletic sock'' attends to $ur3$ -- a previous review on a T-shirt -- the most, and $ur2$ -- a historical review on a Casio watch -- the least. This makes sense since socks share more common properties with T-shirts than watches, such as the material and the fitness. For the item reviews, \modelname{} allocates the most weight to $ir9$ which includes a lot of useful information for ``men big tall active athletic sock'', such as that the socks are for male, they are cushiony, and they stay up on legs.  On the contrary, $ir4$, which receives the least attention, does not reveal whether the socks are for men or women, and the descriptions such as ``comfortable'' and ``'durable'' can also be applied to other products. The attention weights can help explain why \modelname{} ranks this item to the top, and showing reviews with large attention weights to the user could help them better understand why the item is a good candidate and facilitate their purchase decisions. 
%  can effectively identify useful information carried in the reviews. 

\begin{table*}
	\caption{A case of query, user, and purchased item for the multi-layer RTM case study. The attention weights are average from all the 8 attention heads. }
	\centering
	\label{tab:cd_case} 
	%\scriptsize
	\footnotesize
	%   \addtolength{\tabcolsep}{3pt} 
	%    \scalebox{0.95}{    
	\begin{tabular}{C{0.4cm} |C{0.4cm}|C{0.4cm} | p{0.2cm} |p{5.5cm} | C{0.4cm} | C{0.4cm}|C{0.4cm} |C{0.2cm} | p{6cm} }
		\hline
		\multicolumn{10}{c}{Query: \textbf{"CDs vinyl Europe jazz"} (Attention weight w.r.t $q^{(1)}$: 0.097; w.r.t $q^{(0)}$: 0.215)} \\ % (Self-Attention Weight:0.246)} \\
		\hline
		%            \multicolumn{3}{c|}{User Reviews} & \multicolumn{3}{c}{Item Reviews} \\
		%            \hline
		\multicolumn{3}{c|}{Attention w.r.t.} & \multirow{2}{*}[-1.5ex]{ID} & \multirow{2}{*}[-1.5ex]{User Reviews} &
		\multicolumn{3}{c|}{Attention w.r.t.} & \multirow{2}{*}[-1.5ex]{ID} & \multirow{2}{*}[-1.5ex]{Item Reviews}  \\
		\cline{1-3}  \cline{6-8} 
		$q^{(1)}$ &$r_{ui_{u4}}^{(0)}$ & $r_{u_{i3}i}^{(0)}$& & & $q^{(1)}$ &$r_{ui_{u4}}^{(0)}$ & $r_{u_{i3}i}^{(0)}$ & & \\
		\hline
		%        0.4433 &0.5210 & 0.2110 & ur4 & review
		%        &1.5660 &0.9190 & 1.0476 & ir3 & review \\
		%        \hline
		%        0.4236 &0.5221 & 0.2233 & ur8 & review
		%        &1.3369 &0.8827 &1.0411 & ir1 & review\\    
		%        \hline
		%        0.4134 &0.5813 & 0.2270 & ur3 & review
		%        &1.0155 &0.8712 &0.9157 & ir2 & review\\
		
		0.055 & 0.065 & 0.026 &  ur4  &  ... Before buying this I already owned 12 \textbf{Garbarek} albums, must admit though \textbf{I'd pretty much heard "all" he had to offer} ... %However,  a trip overseas lead me to sample a collection which included this cd \& also "Rosensfole" (which, by the way, is also highly recommended). Well, the first track "He comes from the north" just simply enraptured me with its wonderful sprightly ostinato rhythm allowing Garbarek's saxophone to joyfully dance in, around \& right through the piece.
		& 0.196 & 0.115 & 0.131 &  ir3  &  ... I really liked \textbf{Jan} in the seventies.  ... \textbf{I believe it is Jan's best recording in a long time.} I don't prefer it to his earlier Avante guard or \textbf{jazzier} stuff ..., and \textbf{Jan}'s always great solos.     \\ \hline %but this is a long way from Kenny G.
		
		0.053 & 0.065 & 0.028 &  ur8  &  Aural attack To all current \textbf{EST} fans - if news of \textbf{Svensson}'s death wasn't hard enough to take, the music on this album is what you might call "tough love"... %To any of you just curious about \textbf{EST}, this is probably not the album to start your collection as, on the whole, it is neither typical nor representative. %Why?  Well, EST's usual melodic acoustic lyricism is on some tracks embellished, on others supplanted \& on yet others overwhelmed by electronic effects (occasionally quite distorted).  These effects are particularly to the fore on two lengthy tracks one of which "Premonition - Earth" (17min \& which also features a relentless brutal backbeat for the last 5mins) works amazingly well while the other "Leucocyte - Ad Mortem" (13min) sees the effects take over \& grate a little too much for the first half of the piece only to give way to a slightly unsettling "calm" over the second half.
		& 0.167 & 0.110 & 0.130 &  ir1  &  Excellent introduction to \textbf{Garbarek}'s world for newbies ...If you like modern \textbf{jazz} and/or ECM-style musicians, buy it without hesitation! ... % The second CD sounds more like modern \textbf{jazz}/fusion with solos from each group member. Excellent live sound, tight playing of all members of the band, a good set of tunes from different Garbarek's albums.What that guy that left negative review didn't like so much about this album I have no clue!
		\\ \hline
		
		0.052 & 0.073 & 0.028 &  ur3  &  ... there's much more of \textbf{Steve Reich}, \textbf{Christian Wallumrod} \& even \textbf{Esbjorn Svensson} here, meaning that whilst every piece is structured such that you can often predict when an established, largely minimalist pattern is going to change ... %you're often surprised as to exactly how this happens \& where the change leads...  %In virtually all cases, the changes are exciting, engaging \& never one-off (ie. each piece changes instrumentation, mood \& dynamic several times \&, apart from three brief blazing saxophone interludes in Module 45, these are never abrasive).
		& 0.127 & 0.109 & 0.114 &  ir2  &  ... Why some people adore this \textbf{Garbarek}'s release while others simply hate it? ... Not "excellent" but for sure "not bad"... This is - in essence - an IMPROVISATION album. If you are not open to this freestyle, probably you'll not like it, ... Anyway, decide with your own ears and criteria. %I own this album for a couple of years, and I can say it's really good.
		\\ \hline
		
	\end{tabular}
\end{table*}

\begin{figure*}
	\centering
	\begin{subfigure}{.25\textwidth}
		\includegraphics[width=1.7in]{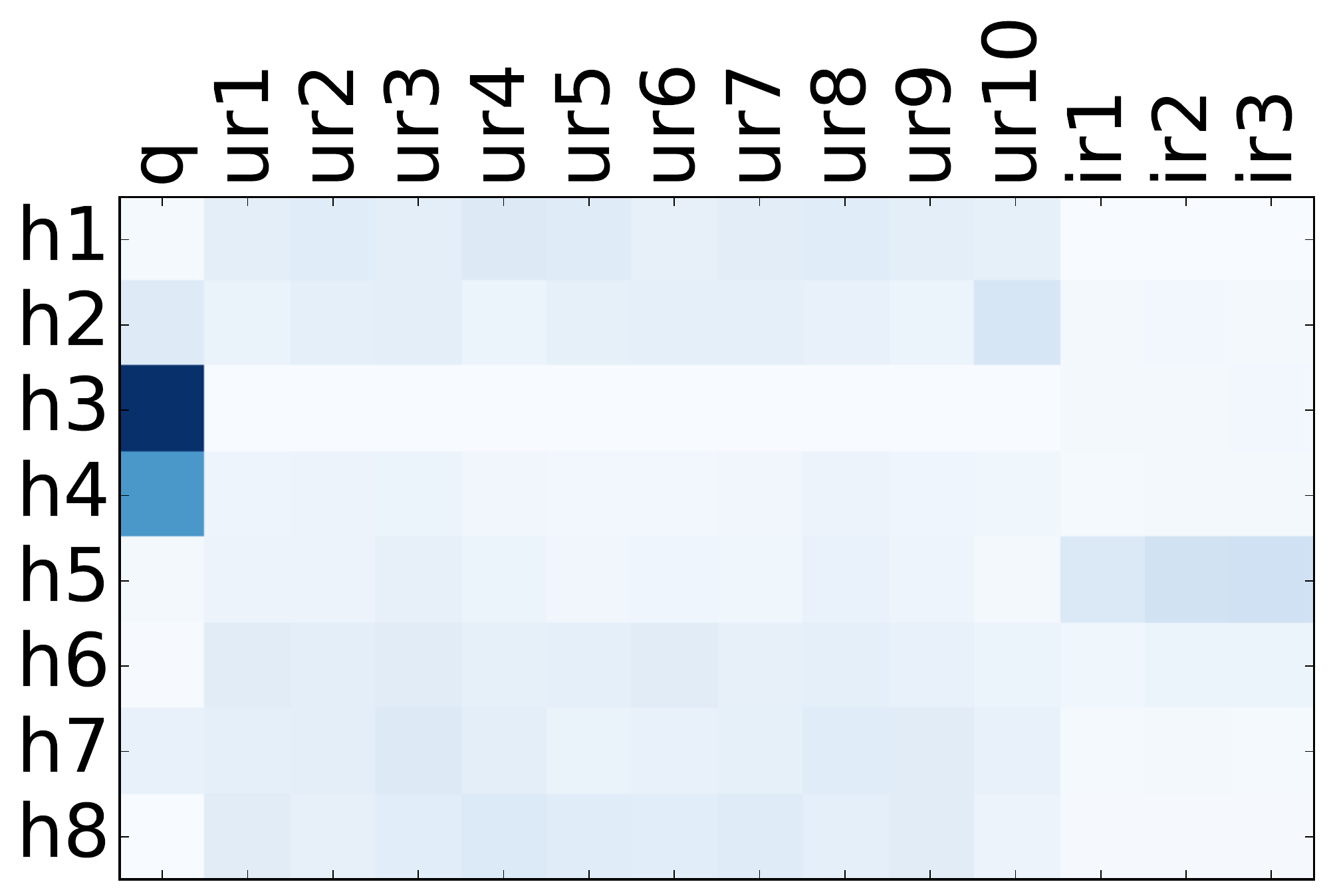}
		%\caption{Percentages of queries influenced by \OurModel with different feedback sizes in \textit{Cell Phones\&Accessories}.}
		\caption{Attention with respect to $q^{(0)}$}
		\label{fig:q0_attn}
	\end{subfigure}%
	\begin{subfigure}{.25\textwidth}
		\includegraphics[width=1.7in]{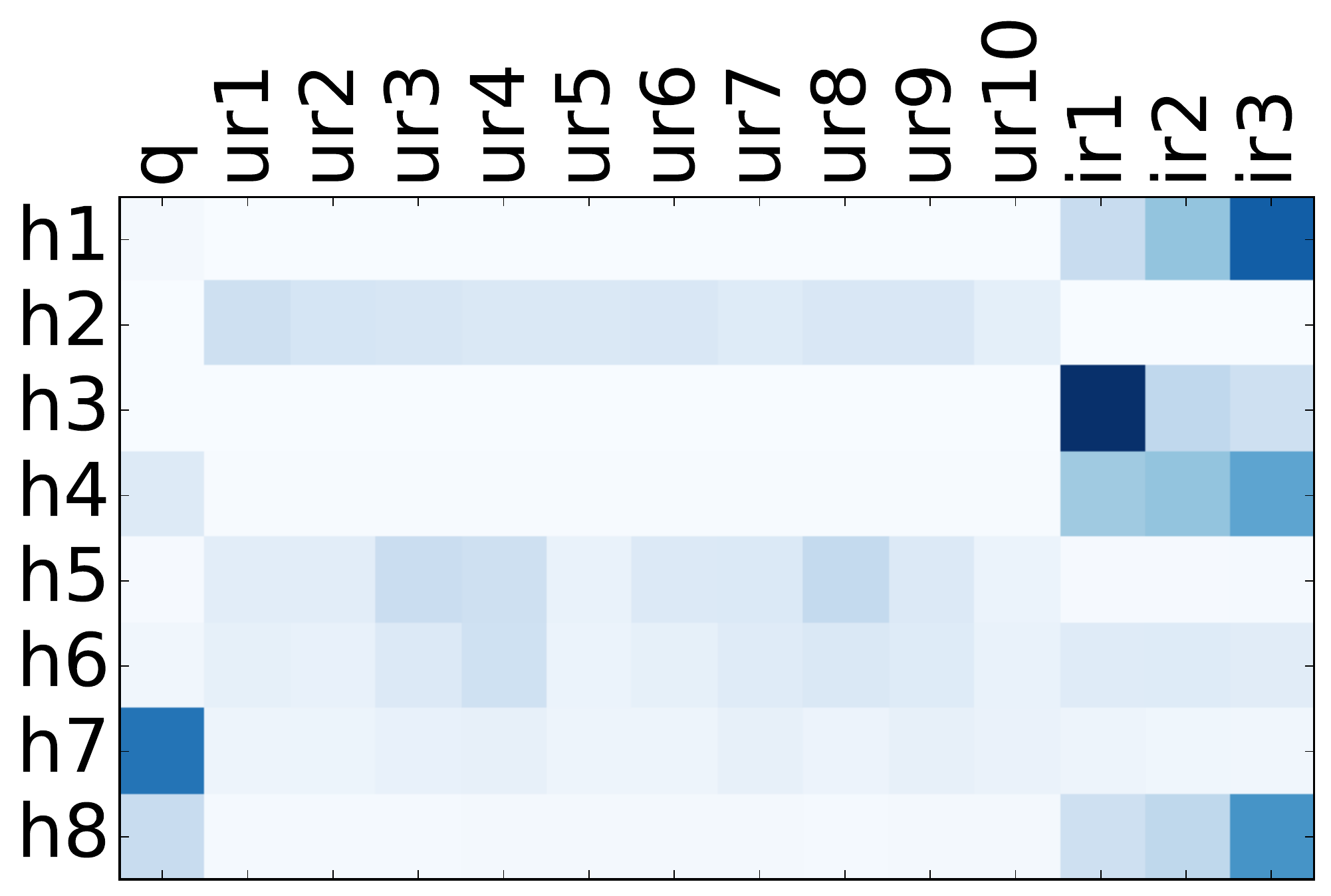}
        \caption{Attention with respect to $q^{(1)}$}
		\label{fig:q1_attn}
	\end{subfigure}%
	\begin{subfigure}{.25\textwidth}
		\includegraphics[width=1.7in]{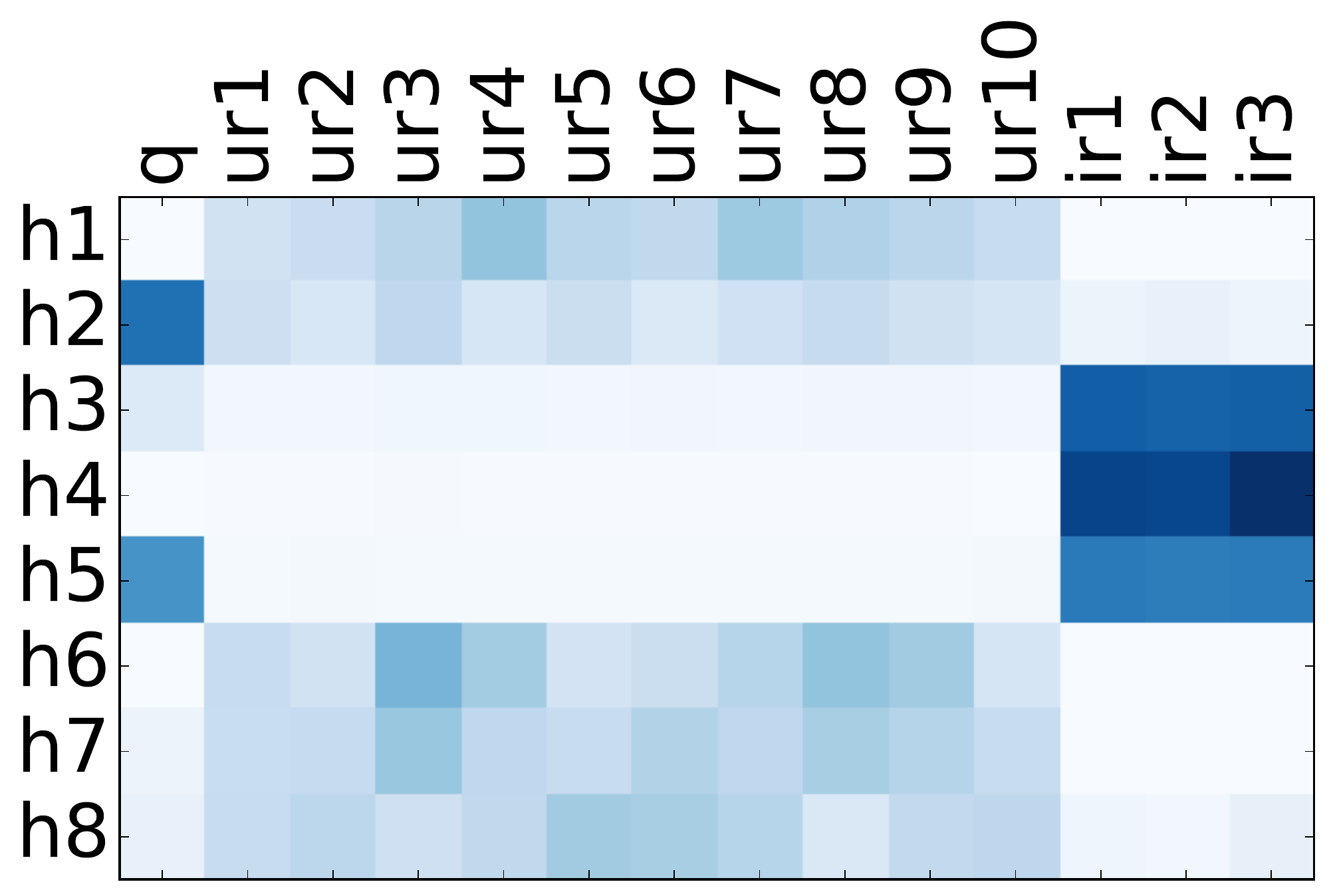}
		\caption{Attention with respect to $r_{ui_{u4}}^{(0)}$}
		\label{fig:ur4_attn}
	\end{subfigure}%
	\begin{subfigure}{.25\textwidth}
		\includegraphics[width=1.7in]{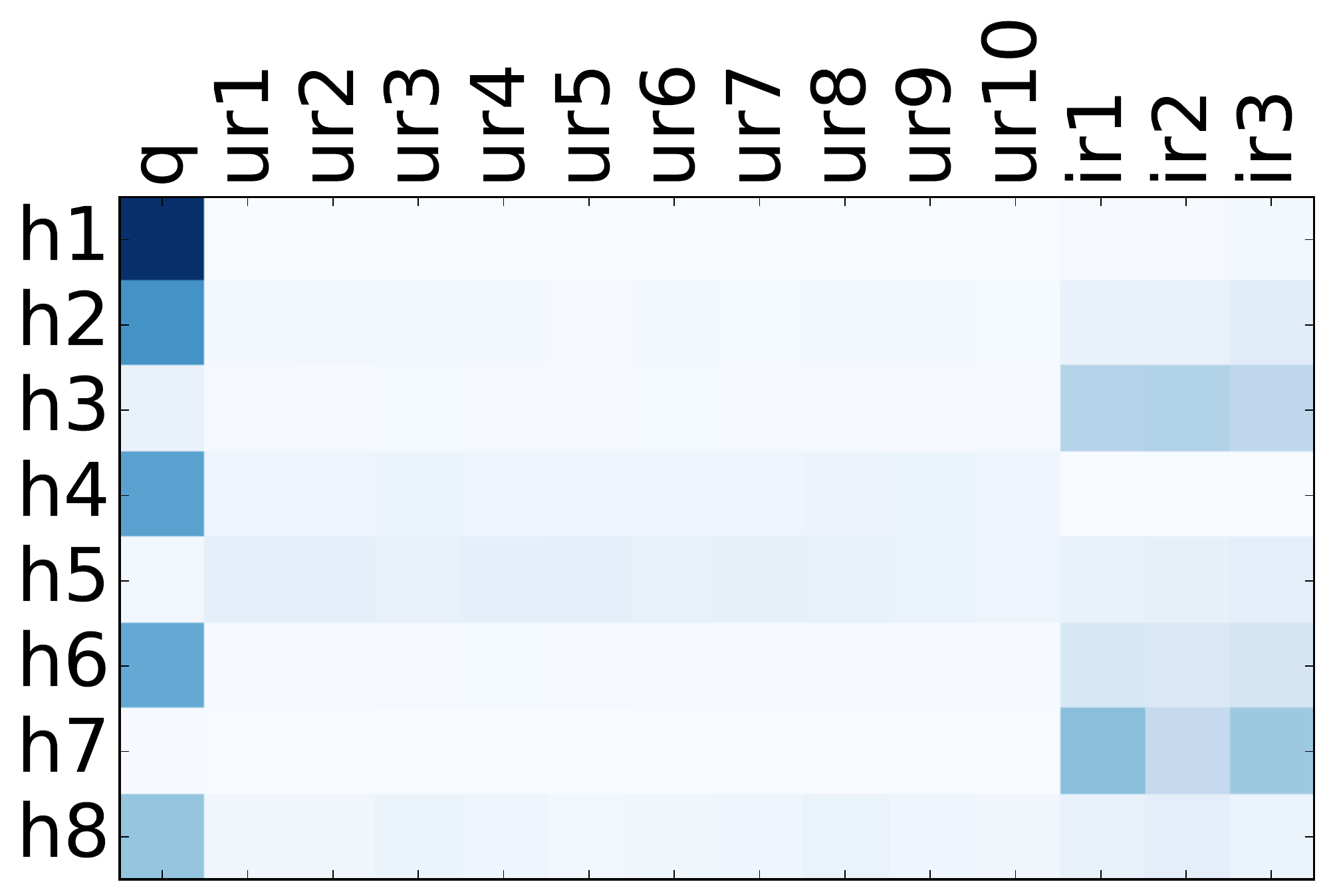}
		\caption{Attention with respect to $r_{u_{i3}i}^{(0)}$}
		\label{fig:ir3_attn}
	\end{subfigure}%
	\caption{Attention weights for the case shown in Table \ref{tab:cd_case}. $r_{ui_{u4}}$ and $r_{u_{i3}i}$ denote the same review as $ur4$ and $ir3$ respectively. }
	\label{fig:cd_attn}
\end{figure*}
\textbf{Case Analysis for Multi-layer \modelname{}}. 
% \textbf{Case in CDs\&Vinyl}. 
%The best \modelname{} on $CDs$ has two transformer layers and add only segment embeddings to the initial unit embeddings (See Equation \ref{eq:q0}). 
For query ``CDs vinyl Europe jazz'', the attention scores of each unit in the sequence of the query, user reviews, and the reviews of a purchased item with respect to different units are shown in Figure \ref{fig:cd_attn}. In the first transformer layer, most of the attention is paid to the query itself by $q^{(0)}$, shown in Figure \ref{fig:q0_attn}. Figure \ref{fig:q1_attn} shows that in the second layer, item reviews draw most attention weights from $q^{(1)}$. These two figures imply that a single layer is not enough to learn the dynamic matching between the query-user pair and the item, probably because the initial representations are not informative enough. 
% With more interactions in the second layer between each unit in the sequence, their representations are dynamically changed according to other units in the sequence. 

From the average attention weight of each unit with respect to $q^{(1)}$ from the 8 heads, the 3 item reviews have the top 3 attention scores, ranked as $ir3,ir1,ir2$ (in Figure \ref{fig:q1_attn}). These attention weights offer a possible explanation for how this item is scored. We can verify the explanation's rationality by checking the text of the item reviews shown in Table \ref{tab:cd_case}. $ir3$ mentions that this CD is Jan's best recording in a long time and Jan Garbarek is a Norwegian jazz saxophonist, which is relevant to the query ``CDs vinyl Europe jazz''. $ir1$ recommends the CD to people who like modern jazz and implies that it is from a European musician - Jan Garbarek. $ir2$ does not mention jazz at all and considers this album not bad and also not excellent. $ir3$ is more positive than $ir1$ on the album, and both of them are more informative and positive than $ir2$, which indicates the explanation of the item score from \modelname{} is reasonable. 

Among the user reviews, $ur4,ur8$, and $ur3$ receive the top attention with respect to $q^{(1)}$ (in Figure \ref{fig:q1_attn}). From the review text shown in Table \ref{tab:cd_case}, we can see that $ur4$ is written by the user for a previously purchased album from the European jazz musician Jan Garbarek. In $ur8$, EST is short for ``Esbj\"orn Svensson Trio'', which was a Swedish jazz piano trio, indicating the album with the review is also related to European jazz. $ur3$ mention Steve Reich (an American composer who also plays jazz), Christian Wallumrod (a Swedish jazz pianist), and Esbj\"orn Svensson (a Norwegian jazz musician), which are also related to the query. These reviews are useful to indicate the user's preference for European jazz and should draw more attention with respect to the query. 

To find out which unit has more effect on $ur4$ and $ir3$ in the first transformer layer, we also show the attention distribution of each unit with respect to $r_{ui_{u4}}^{(0)}$ (ur4) and $r_{u_{i3}i}^{(0)}$ (ir3) in Figure \ref{fig:ur4_attn} and \ref{fig:ir3_attn}. For $ur4$, the top 5 units that it attends to are $ir3$, $ir1$, $ir2$, $ur3$, $q$, and $ur8$, from which we can tell that $ir3$ has the largest contribution in terms of satisfying preferences indicated by $ur4$. As shown in Table \ref{tab:cd_case}, $ur4$ indicates that the user is a big fan of Jan Garbarek and has listened to almost all his albums, so the comment in $ir3$ that this album is the best of Garbarek is quite persuasive to the user to purchase the item. 
%The other parts that affect the dynamic representation of $ur4$ are $ur3$, $q$, and $ur8$, which are related to $ur4$ and could help adjust its embedding vector. 
In Figure \ref{fig:ir3_attn}, the units with top attention weights with respect to $ir3$ are $q, ir3, ir1, ir2, ur3, ur8, \text{and }ur4$. This implies that the final representation of $ir3$ depends mostly on the query and all the item reviews, including itself. The attention weights also indicate that $ir3$ considers itself to satisfy the query and the preferences expressed in the user reviews $ur3, ur8, \text{and }ur4$ the most. 
These observations are consistent with our interpretation on the dynamic review representation of a multi-layer \modelname{} in Section \ref{subsec:model_interpretation}.

	%!TEX root=main.tex
\begin{comment}
Explainable search. 
comparable attention scores, explanation for personalized product search. (which historical item draws more attention; whether it is necessary to be personalized. )
Other ways of incorporating transformer that consider other information of products. 
%More analysis based on query characteristics. 
\end{comment}
\section{Conclusion and Future Work}
\label{sec:conclusion}
In this paper, we propose a review-based transformer model (\modelname{}) for personalized product search, which scores an item by encoding the sequence of the query, user reviews, and item reviews with a transformer architecture. \modelname{} conducts review-level matching between a query-user pair and an item instead of the popular paradigm that represents and matches users and items with static representations in the semantic space. Each user and item review's importance is dynamically changed according to other units in the sequence, which enables \modelname{} to perform adaptive personalization and the dynamic utilization of the item information in different search sessions. 
%The attention weights of reviews indicate their importance during item scoring. 
The empirical results show that \modelname{} not only improves product search quality but also provides useful information to explain why an item is ranked to the top. 

As a next step, we are interested in investigating incorporating other information from users and items such as brand, category, the relationship of also-purchased, and also-viewed, etc., with transformers to weigh each information source dynamically across search sessions. 
%We also plan to find a way to provide explanations with the attention weights collected in different transformers layers from all the attention heads. 
%User studies would help us confirm whether the explanations are reasonable and helpful. 

\begin{acks}
	This work was supported in part by the Center for Intelligent Information Retrieval. Any opinions, findings and conclusions 
	or recommendations expressed in this material are those of the authors and do not necessarily reflect those of the sponsor.
\end{acks}

	\bibliographystyle{ACM-Reference-Format}
	
	\bibliography{reference}
	
\end{document}